\documentclass[paper]{sigirforum}

\def\pubissue{Vol. 56 No. 2 December 2022}

\usepackage{natbib}
\usepackage{booktabs}

\begin{document}
\title{Data Management Challenges for Internet-Scale 3D Search Engines}

\authors{
\author[jamesbw@cs.toronto.edu]{James Williams}{University of Toronto}{Canada}
\and
\author[sscott@physna.com]{Shane Scott}{Physna Inc.}{USA}
\and
\author[swedig@physna.com]{Sean Wedig}{Physna Inc.}{USA}
\and
\author[thindanov@physna.com]{Timur Hindanov}{Physna Inc.}{USA}
\and
\author[croedig@physna.com]{Christoph Roedig}{Physna Inc.}{USA}
}

\maketitle 
\begin{abstract}
This paper describes the most significant data-related challenges involved in building internet-scale 3D search engines. The discussion centers on the most pressing data management issues in this domain, including model acquisition, support for multiple file formats, asset versioning, data integrity errors, the data lifecycle, intellectual property, and the legality of web crawling. The paper also discusses numerous issues that fall under the rubric of trustworthy computing, including privacy, security, inappropriate content, and copying/remixing of assets. The goal of the paper is to provide an overview of these general issues, illustrated by empirical data drawn from the internet's largest operational search engine. While numerous works have been published on 3D information retrieval, this paper is the first to discuss the real-world challenges that arise in building practical search engines at scale.
\end{abstract}



\section{Introduction}\label{sec_intro}

\textit{Three-dimensional} (``3D'') modeling is increasingly common in the modern world, playing a prominent role in such diverse industries as motion pictures, video games, medical research, architecture, and industrial engineering. As the amount of 3D content grows, there is a corresponding need for practical information retrieval tools. 

This paper discusses the most pressing data-related challenges facing internet-scale search engines for 3D models. In contrast to text-based search engines like Google, Bing, or Yandex, this type of information retrieval system allows users to search the internet for 3D assets by geometry as well as by text. Unlike the prototypes described in the literature, which typically use small databases consisting of tens of thousands of models, an internet-scale search engine will accumulate datasets in the tens or hundreds of millions.

\newpage 
Although there are many  approaches to 3D information retrieval, any real-world search engine will be faced with several common data management issues, including model acquisition, diversity of file formats,  intellectual property, the legality of web crawling, and trustworthy computing (e.g., privacy, security, objectionable content). On the engineering side, data lifecycles,  asset versioning, and data integrity are considerations that cut across the choice of 3D search algorithm.

\begin{figure}[ht]
    \centering
    \includegraphics[width=6in]{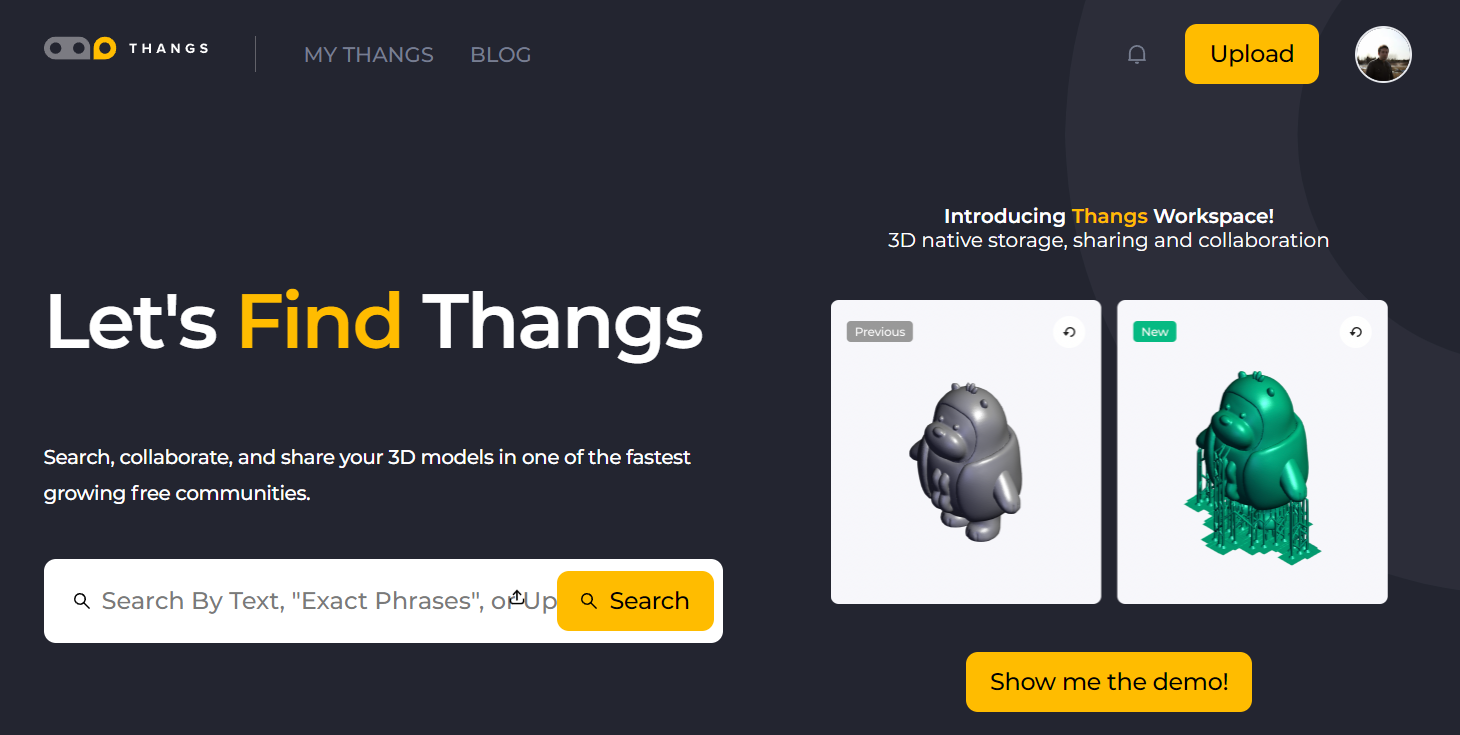}
    \caption{The landing page of the Thangs 3D search engine.}
    \label{fig:thangs}
\end{figure}

These issues are among the many challenges encountered by the  authors during the design and development of the largest 3D search engine on the internet. Owned and operated by Physna Inc.,  \textit{Thangs.com} (``Thangs'')  is both an internet-scale search engine and a content hosting site where users can upload models (see Figure \ref{fig:thangs}).  

At the time of writing, Thangs provides access to 14,800,000  models through text-based and geometric search. Most of these models are geared towards 3D printing and computer games, but the number of engineering-oriented artifacts is growing steadily. Around 300,000 models were uploaded by users through the platform's content hosting functionality, while the rest were acquired by web crawling.

\begin{figure}[ht]
    \centering
    \includegraphics[width=5.0in]{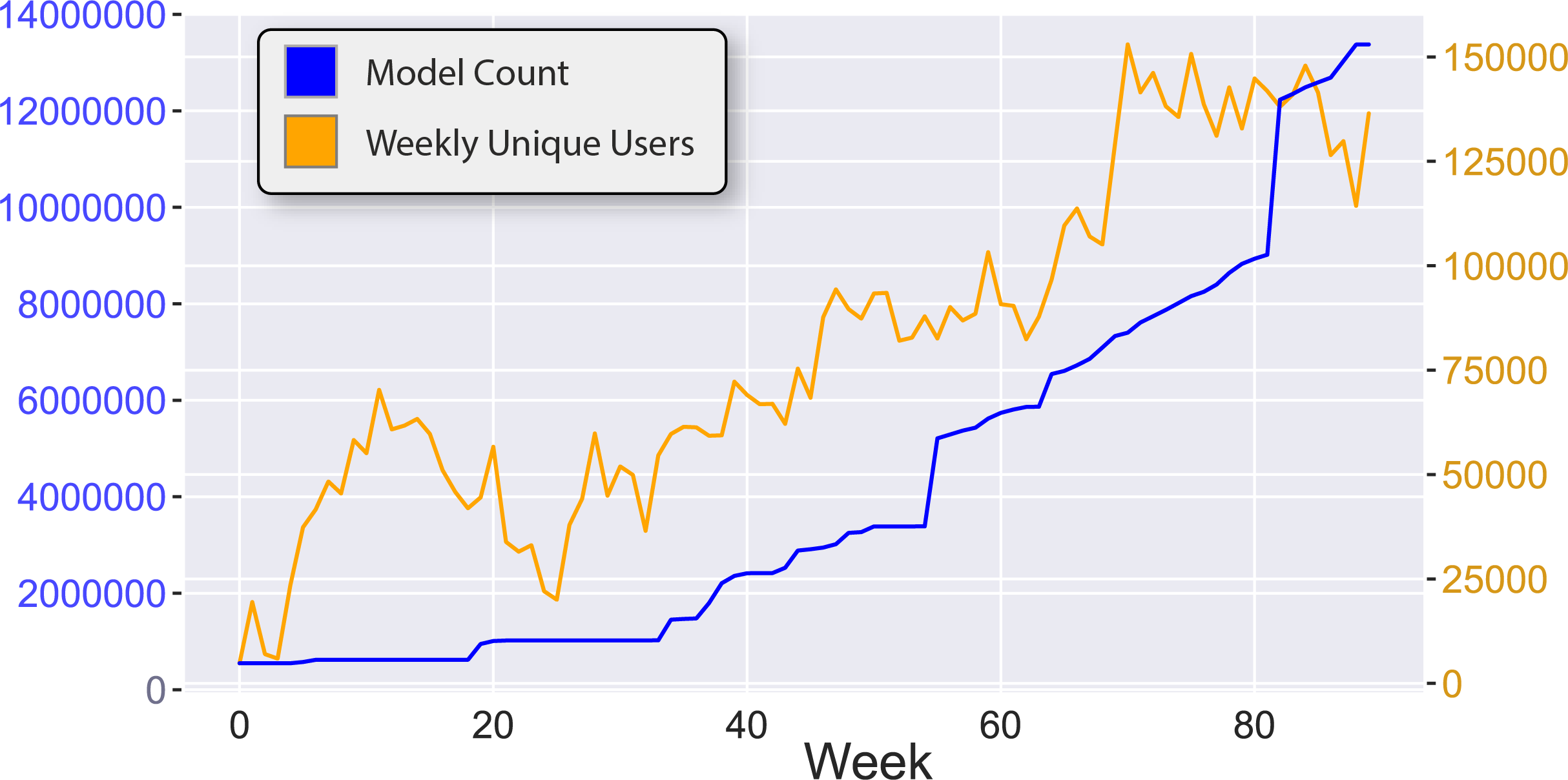}
    \caption{Model count (blue) and weekly unique users (orange) by week of operation, beginning in August of 2020.}
    \label{fig:model_count_and_user_count}
\end{figure}

The performance of the search engine is tracked by 98 weekly metrics, including: (1) unique users; (2) session duration; (3) number of text searches; (4) number of models acquired through web crawling, and; (5) number of model searches performed by  users. Figure \ref{fig:model_count_and_user_count} shows weekly unique users (orange) and the number of models (blue) from its release date in August 2020 to the first draft of this paper in May 2022. For comparison, data concerning Thingiverse.com can be found in recent works (e.g., \citep{Flath2017}).

Since the Thangs search engine is operated by a early-stage startup company  there are obvious limitations on what the authors can reveal.  Nevertheless, the paper has at least two merits: (1) it contains a novel discussion of the key data management issues that will confront any internet-scale, 3D search engine, and; (2) it presents empirical data drawn from the world's largest operational 3D search platform. Certain problems become more pressing when a database grows to petabyte scale.

The structure of the paper is as follows. Section 2 provides background information, organized by topic. Section 3 discusses challenges accompanying data collection, including file format types and legal risks associated with web scraping. Section 4 surveys relevant intellectual property law from the United States, while Section 5 deals with trustworthy computing issues---namely, privacy, security, inappropriate content, and copying/remixing of model data. Section 6 covers some key challenges facing 3D information retrieval schemes, with a particular focus on ``bag of words'' approaches that rely on local neighborhood measurements. Section 7 briefly describes the use of training and test datasets. Finally, Section 8 discusses data lifecycle issues. The paper concludes with suggestions for future research.

\section{Background}\label{sec_background}

The discipline of \textit{information retrieval} (``IR'') focuses on representing, searching, and manipulating collections of useful information \citep{Buttcher2010}. This paper is concerned with 3D IR (i.e., shape retrieval), which has been identified as vital for such diverse tasks as digital archiving and  manufacturing process optimization \citep{Huang2015}. Economists have suggested that this type of technology is a key component of the ``fourth industrial revolution'' \citep{Xu2018}.

The authors of this paper were unable to find research publications that discuss data management issues for large-scale 3D search engines. However, previous works (i.e., \citep{Corney2002,Rea2002}) touch upon similar themes in the course of demonstrating a small, online 3D search engine. Some of the design decisions made in those works are in keeping with the Thangs search engine, including the use of: (1) triangular meshes as a fundamental primitive; (2) multi-stage filtering, and; (3) global shape descriptors. However, the search engine databases in those works appear to contain hundreds of models rather than the tens of millions supported by Thangs. Furthermore, the authors mention web crawling in passing but do not discuss any of the significant challenges facing data acquisition.

Since this paper covers many different issues pertaining to 3D search engines, it is necessary to summarize the literature on a per-topic basis. The succeeding sections cover previous works in the following order: (1) information retrieval for 3D models; (2) the regulation of web crawling; (3) intellectual property; (4) remixing of 3D models, and; (5) attacks using 3D assets.

\subsection{Information Retrieval for 3D Models}

A large body of research exists on the topic of 3D IR. Among the most important search modalities are: (1) \textit{exact match search} (``EXS''); (2) \textit{similarity search}, also known as \textit{3D shape search} (``3DSS'') \citep{Iyer2005}, and; (3) \textit{part-in-part } (``PIP'') search, also known as ``part-in-whole'' (``PIW'') retrieval  \citep{Muraleedharan2019}. The Thangs search engine supports all three search variants, as well as standard \textit{text-based search} (``TBS''). Physna's enterprise products are the core of its business, and PIP is the most important search type for the company as a whole. 

%
\subsubsection{3D Similarity Search} As described in Section \ref{sec:bow_retrieval}, there are many representations of 3D objects and also many ways to define similarity relationships between them. A large number of methods have been proposed for 3DSS, including locality-constrained sparse patch coding (e.g., \citep{Liu2015}) and convolutional neural networks (e.g., \citep{Wang2019}). For purposes of this paper, the most relevant of these are:
\begin{enumerate}
  \setlength\itemsep{0.5em}
    \item view-based approaches (e.g., \citep{Zhang2019}), and;
    \item ``\textit{bag of features}'' (``BOF'') methods (e.g., \citep{Csurka04visualcategorization,Liu06shapetopics}).
\end{enumerate}
A detailed review of the literature is beyond the scope of this paper, but it is worth mentioning two examples that combine these techniques.

\citet{MASOUMI2017198} introduced a 3DSS method that incorporates shape descriptors at multiple levels: (1) local descriptors are generated by a spectral graph wavelet transform; (2) mid-level features are obtained by embedding the local descriptors into a visual vocabulary space, and; (3) global descriptors are constructed by aggregating the mid-level features. This approach provides an excellent example of how to use sets of descriptors at different scales. Thangs uses a multi-stage filtering scheme that makes use of local, intermediate, and global descriptors; however, the mid-level descriptors are used in a different way than in Masoumi and Hamza's work.

\citet{Ohbuchi2008} developed a hybrid approach in which: (1) images are generated by rendering views of a 3D model at different viewpoints; (2) local features are extracted from the images; (3) vector quantization is applied to the features to generate a bag of ``visual words'', and; (4) a histogram is constructed for each bag, serving as the model's feature vector. This approach integrates view-based and BOF approaches in an entirely different way than the Thangs search engine, which uses a view-based technique for 3DSS and a bag-of-words technique for EXS and PIP. Generating embeddings (or signatures) of 3D models is a standard technique in the literature (e.g., \citep{Tangelder2008,Ip2002,Osada2001}), and one that is also used by Thangs for 3DSS. However, Thangs deliberately avoids quantization of local measurements in other contexts due to the issues that arise in large datasets (see Section \ref{sec:collisions}).


\newpage 

%
\subsubsection{Part-in-part Search} 

PIP search methods allow a user to find models in a database that contain a query object as a part \citep{Muraleedharan2019}. This is a particularly compelling use case for corporate entities that maintain large datasets of 3D models. There are a variety of formulations of PIP, including partial retrieval. A recent survey of the field was presented in \citet{Liu2013}, who partition existing methods into three categories: (1) local descriptor methods, which analyze part-in-part relationships using local neighborhood measurements; (2) segmentation methods, which divide a 3D object into parts through segmentation, and; (3) view-based methods, which generate a set of 2D images from a 3D model. The PIP method found in Thangs is a multi-stage filtering approach based on local descriptors.



\subsection{Regulation of Web Crawling}

The legality of web crawling is a topic of increasing interest in the legal literature, as evidenced by a growing number of journal articles (e.g., \citep{Christensen2020,Drivas2019}). However, these works largely focus on acquisition of personal information (e.g., social media profiles) as opposed to commercial artifacts like 3D models. For instance, a recent paper by  \citet{Sobel2021} provides an analysis of web scraping under United States law, contending that tort law may provide a deterrent to the scraping of personal data.

One significant drawback in relying upon the secondary research literature for information on current regulations is that these works can become outdated in short order. One of the major cases clarifying federal regulation of web scraping in the United States \citep{HiQ2022} was decided between drafts of this article; that decision made some of the most pertinent works in the secondary literature (e.g., \citep{Yannella2021,Christensen2020}) outdated on that topic. As a result, it is mandatory to perform proper legal research by using the secondary literature only as a guide to important primary sources.

It is easy to find technical books that cover the software engineering aspects of web crawling (e.g., \citep{Jarmul2017}). While some of these works discuss legal issues (e.g., \citep{Mitchell2018}), it is a bad idea to rely upon the software engineering literature for legal analysis. A recent book by \citet{Patel2020} discusses web scraping at scale, but 3D search engines are not primarily limited by computing resources---rather, they face significant challenges in gaining access to data in the first place.

\subsection{Data Integrity Issues with 3D Models}

3D models encountered on the internet are frequently afflicted by data integrity errors, including flipped normals, non-manifold edges, degenerate triangles (e.g., those with two identical vertices), and duplicated triangles (see, for example, \citep{DBLP:journals/corr/ZhouJ16}). Unlike academic research projects that rely on curated datasets (e.g., the Amazon/Berkeley Object Dataset \citep{Collins_2022_CVPR}), a 3D search engine cannot assume that 3D artifacts are error-free. The Thangs search engine routinely encounters 3D models that cannot be handled by popular libraries (e.g., Open3D \citep{open3d}) due to various deficiencies.

\newpage
Overviews of data integrity errors and algorithms for repairing them are presented in several articles (e.g., \citep{Attene2013,Ju2009}). The authors of the present work could not find publications that discuss data integrity errors in the context of 3D search engines, but it is clear that data integrity errors are a problem for many search methods. Internet-based 3D search engines are in a difficult position, however, as their goal is to index models in their natural state. Unlike a commercial asset management system for additive manufacturing, a search engine is not tasked with providing quality guarantees; only those data integrity errors that interfere with indexing should be ameliorated.

\subsection{Intellectual Property (``IP'')}

The authors of the current work could not find articles in the research literature that discuss IP-related challenges for 3D search engines. While there are numerous court decisions on trademark and copyright in the context of text-based search engines (e.g., \citep{Authors2015}), including those that  use digital thumbnail/preview images, it is not clear that these cases offer sufficient guidance for a search engine that manages a large collection of 3D models.

Copyright protection for 3D assets has received a significant amount of interest in the secondary legal literature (e.g., \citep{Meurer1997,Flank2017,Vogel2016}). For instance, a relatively recent work by \citet{Dasari2013} discusses the status of 3D models and their corresponding physical objects under United States law; among other topics, the article describes copyright infringement action taken against Thingiverse under the \textit{Digital Millennium Copyright Act} (``DMCA''). A journal article by \citet{Rideout2011} takes a closer look at the status of 3D models under copyright law, as well as the ``safe harbor'' provisions of the DMCA that can be invoked by service providers (e.g., 3D content hosting platforms, 3D search engines) to avoid liability. 

There is also a large body of work in the computing literature that describes techniques for supporting copyright in digital works. Many of these research efforts are focused on developing novel approaches to watermarking and steganography \citep{Cox2008}. Copyright protection for 3D models is well represented. A recent survey of security safeguards for 3D printing is provided by \citet{Hou2018}, who evaluate current techniques (e.g., cryptography, digital rights management, watermarking) against a set of copyright infringement scenarios. Unfortunately, their paper only makes passing reference to the issues of moral rights and derivative works.

A wide variety of watermarking schemes have been developed for 3D models (e.g., \citep{Zafeiriou2005}).  Watermarking schemes must be robust against various attacks, including remeshing, Gaussian noise addition, and mesh simplification \citep{Wang2008}. Watermarking can have an impact on information retrieval systems, since modifications to 3D objects can cause problems for many similarity algorithms. For example, perturbation of mesh vertices can affect the local measurements used in facet-based ``bag of words'' approaches (e.g., \citep{Warner2017}). 

\newpage
\subsection{Remixing of 3D Models}
\label{sec:background_remixing}

The ``remixing'' of multimedia artifacts has been a topic of interest in the humanities and fine/performing arts since the 1970s. Several recent volumes provide an overview of work in the interdisciplinary field known as  ``Remix Studies'' (e.g., \citep{Navas2014,Navas2021}). Remixing and its role in innovation has also been a topic of interest in economics and managerial science dating back to early work by Schumpeter \citep{Schumpeter1942}. 
    
It appears that the computing community took an interest in remixing around a decade ago, and several research teams have published studies that analyze data from popular 3D modeling websites (e.g., \citep{Wirth2015PatternsOR,Voigt2018,Flath2017}). For instance, \citet{Papadimitriou2015} used a network-science approach to analyze remixing of models on Thingiverse. The existing studies indicate that remixing is an important feature of online 3D modeling communities. However, not all users are interested in seeing their content reused, and sites like Thangs and Thingiverse allow users to specify their own license terms. A search engine that provides links to models that either violate license terms (e.g., attribution) or infringe copyright can expect to receive numerous ``takedown'' requests.

\subsection{Attacks using 3D Models}
\label{sec:background_attacks}

Security issues pertaining to 3D models have been examined from a variety of perspectives for at least two decades (e.g., \citep{Harte2002}). For instance, the topic has been studied as a key element of cybersecurity in \textit{additive manufacturing} (``AM'') systems (e.g., \citep{Yampolskiy2016,YanambakaVenkata2020,Sturm2014}). A recent review of papers from the literature was provided by \citet{Yampolskiy2018}, who divide attacks into two high level categories: (1) AM sabotage, and; (2) theft of technical data. \citet{Zeltmann2016} study two structural integrity attacks---one that adds small defects into the interior of a 3D model, and another that re-positions parts on a build plate. \citet{Etigowni2021} provide an interesting, image-based detection system meant to address the first of these attacks, while Straub \citep{Straub2017} examines the second.

Many of the attacks discussed in the literature on additive manufacturing systems are of marginal concern for a 3D search engine. It is not the job of a search engine to guarantee that all models are printable or free from defects, although warning of these quality concerns could be a useful feature. Search engines are concerned with indexing content available on the internet, whether that content has latent defects or not.

Nevertheless, there are several types of model-based attack that are highly relevant to 3D search engines. Attackers can impair the quality and efficiency of a retrieval system by using carefully crafted attack vectors. \citet{Sturm2017} provide a list of attacks against STL files that are relevant to 3D information retrieval schemes, including vertex movement and creation of small indents/protrusions. Further work is needed on the topic of security vulnerabilities in 3D search engines.

\newpage
\section{Data Collection}

Efficient and comprehensive acquisition of data is of paramount importance to any internet search engine.  3D models, however, are typically proprietary \citep{Kim2020}. Search engines are not only confronted by access control mechanisms, but also a tangled web of regulatory constraints that limit both acquisition and subsequent use of 3D model data. This section contains a discussion of the major issues in this domain, including data sources, file formats, versioning, and restrictions on web crawling.

\subsection{Data Sources}
\label{sec:data_sources}

Internet-accessible sources of 3D data can partitioned into four main types: (1) \textbf{public websites} that are designed to provide open access to a database of 3D models (e.g., Thingiverse); (2) \textbf{proprietary websites}, typically owned by distributors or \textit{original equipment manufacturers} (``OEMs''); (3) \textbf{social media websites}, which often contain models that are posted by users to their personal accounts, and; (4) the ``\textbf{dark web}'' \citep{Beckstrom2019}. Game engines (e.g., Unreal) and video games modding communities are a rich source of 3D models, but their content is generally not directly accessible  from a hyperlink.

Apart from public websites, each of these sources of data has disadvantages. Proprietary sites often utilize access control mechanisms that make anonymous web scraping impossible. While social media sites have a significant amount of open access content, there are several issues that make them less useful as a source of models. First, 3D artifacts on social media sites are often ephemeral and consequently not worth indexing. Second, the costs involved in performing a search of these sites are substantial. Finally, models on the dark web are typically accompanied by significant intellectual property concerns. As a result of these considerations, the Thangs webcrawler targets public and proprietary repositories only.

\begin{figure}[ht]
    \centering
    \includegraphics[width=4.5in]{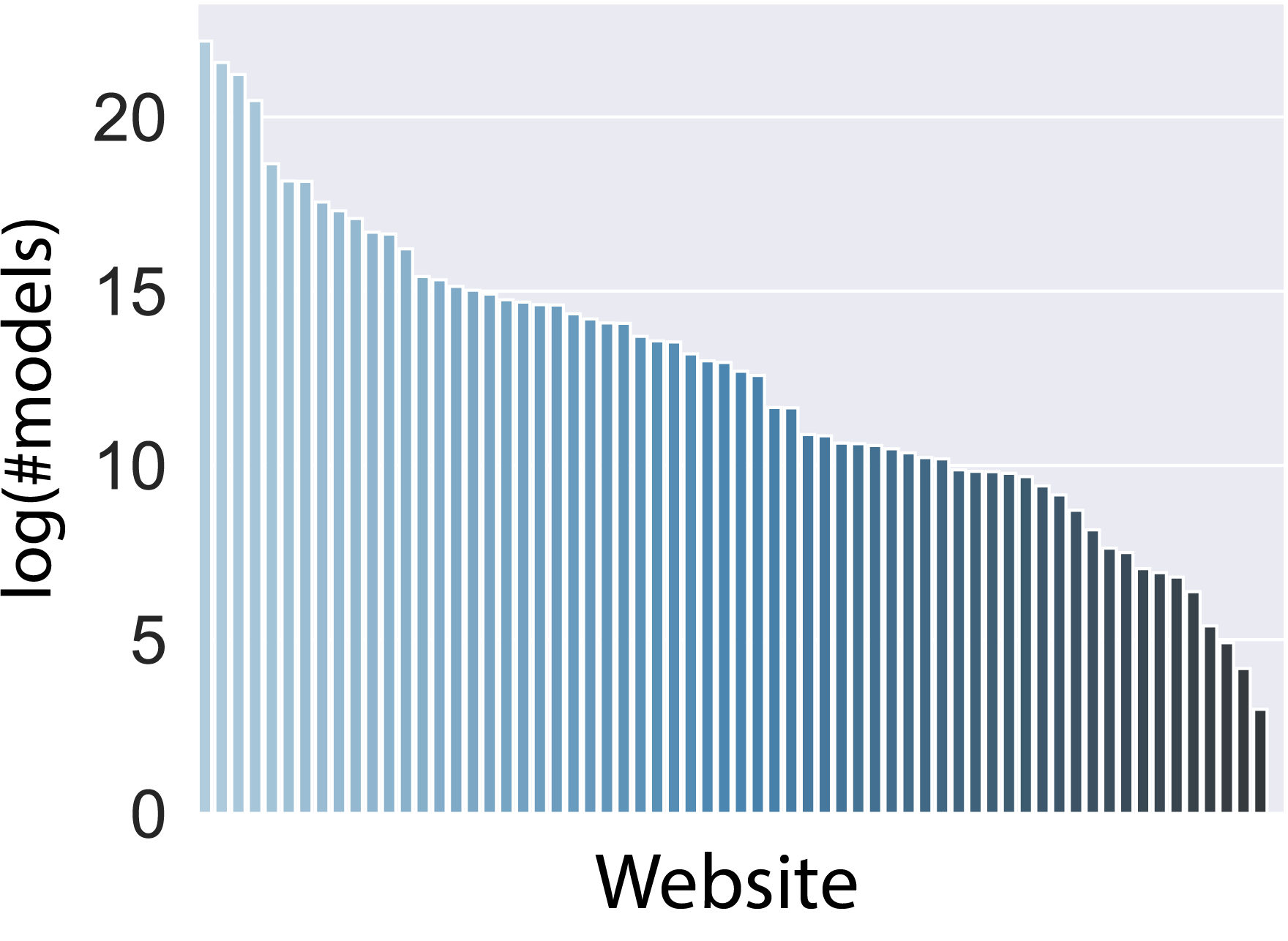}
    \caption{Number of models in the top 70 websites in decreasing order (logarithmic scale).}
    \label{fig:distr_model_counts_per_site}
\end{figure}

Websites follow a ``long tail'' distribution in terms of the number of models that they contain. Figure \ref{fig:distr_model_counts_per_site} shows model counts (in log scale) for the top 70 websites indexed by the Thangs search engine. A handful of websites provide the bulk of the models---for instance, Thingiverse currently hosts around 4.7 million individual 3D model files, with a heavy emphasis on 3D printing artifacts. From this initial group of model-rich websites, model count drops precipitously to the point where many websites contribute mere dozens of models each. 

\subsection{File Formats}
\label{sec:file_formats}
Many different file formats have been created for 3D models  \citep{McHenry2008}. For instance, models intended for use in additive manufacturing (i.e., 3D printing) are invariably saved in the popular STL file format \citep{Szilvasi-Nagy2003} that represents 3D objects as triangle meshes.  Wavefront's OBJ is a mesh-based format that is widely used in video games due to its support for materials and textures. Some file formats, like Autodesk Filmbot's FBX, can store animation and rigging information in addition to 3D geometry. File formats that support \textit{computer-aided design} (``CAD'') tasks are based on proprietary constructive geometry data that preserves the design history of the model; an example of this is the SLDPRT file format used by SolidWorks \citep{Lombard2019}. Formats such as STEP or IGES represent solid bodies with geometric primitives that provide a high degree of accuracy, making them useful for exchanging models between CAD platforms.

Diversity of file formats can be a significant challenge for geometric search, as many techniques for shape similarity assume that datasets contain artifacts in a single, canonical file format.  A search engine could restrict itself to handling a small number of popular file formats. However, the Thangs search engine is also a content hosting system that allows users to upload their own models; supporting a small set of file formats would be a major drawback in terms of usability and uptake.

\begin{figure}[ht]
    \centering
    \includegraphics[width=4.8in]{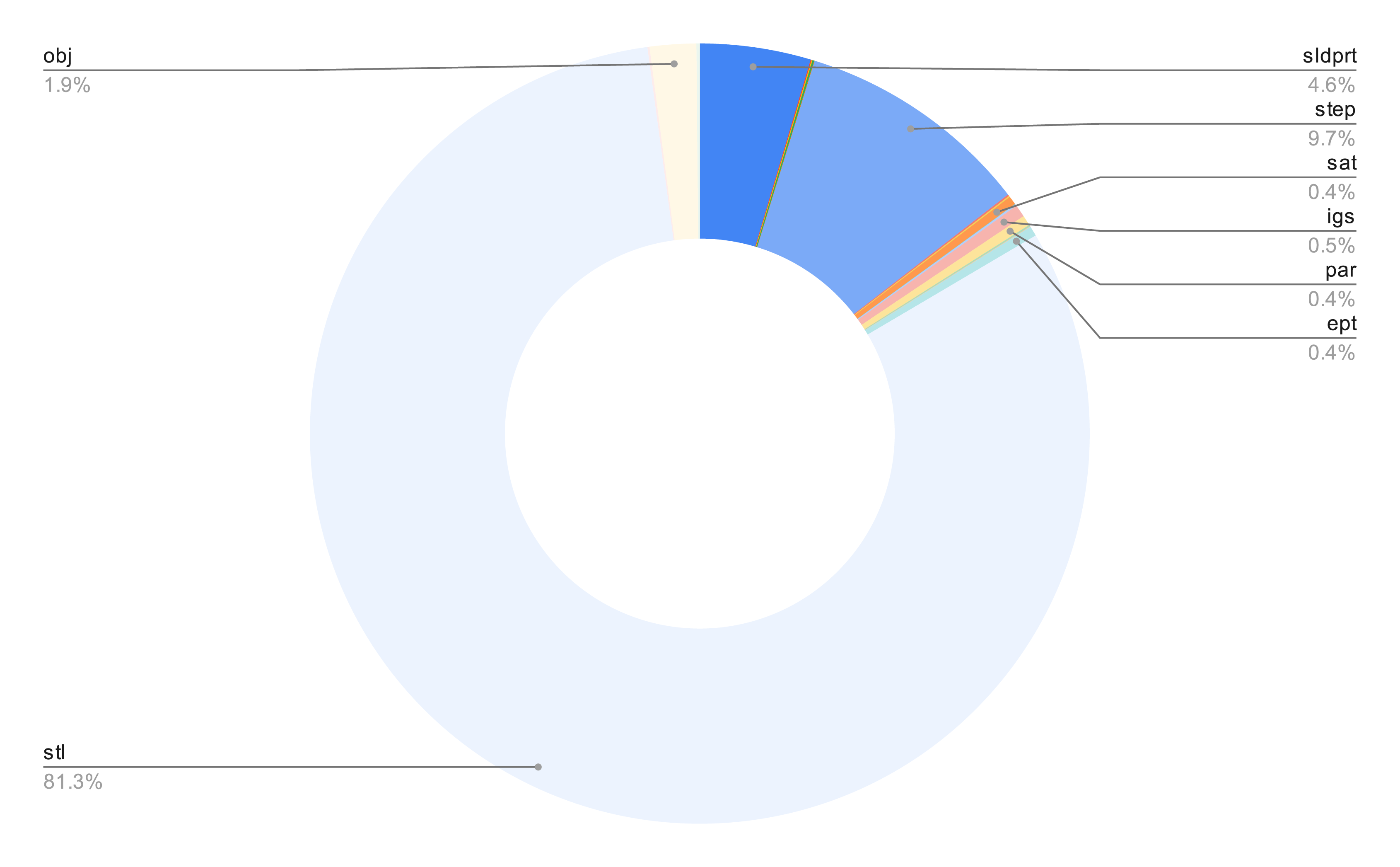}
    \caption{File formats in a sample of 7 million models from the Thangs dataset.}
    \label{fig:file_format_dist}
\end{figure}

Physna handles diverse file formats by converting 3D models into a canonical triangle mesh format with the use of a third party file-conversion library (see also \citep{Kim2020}). This choice affects its web crawling strategy as well---Physna's web crawlers will choose STL over other file formats, if given the option. This preference for STL files is reflected in Figure \ref{fig:file_format_dist}, which shows file formats for a dataset of 7 million models. About 2 million models  represented in STL form are available from OEM sites in other formats, with STEP, SLDPRT, and OBJ being common alternatives. Models intended for 3D printing are usually available only in STL format.

\begin{figure}[t]
     \centering
    \includegraphics[width=4in]{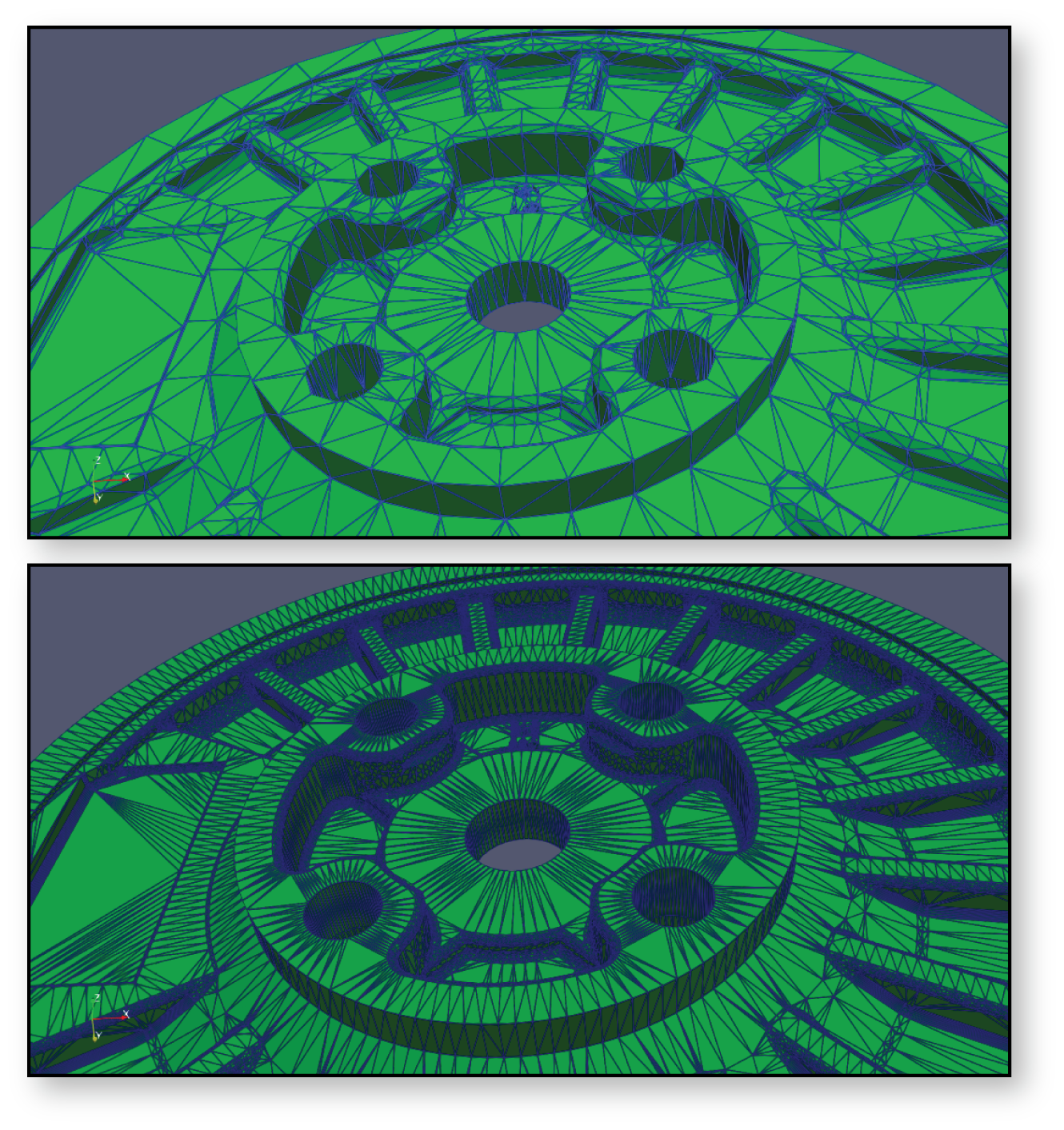}
    \caption{Different tessellations of the same CAD model.}
    \label{fig:tesselation_diff}
    \end{figure}

\begin{figure}[t]
    \centering
    \includegraphics[width=4in]{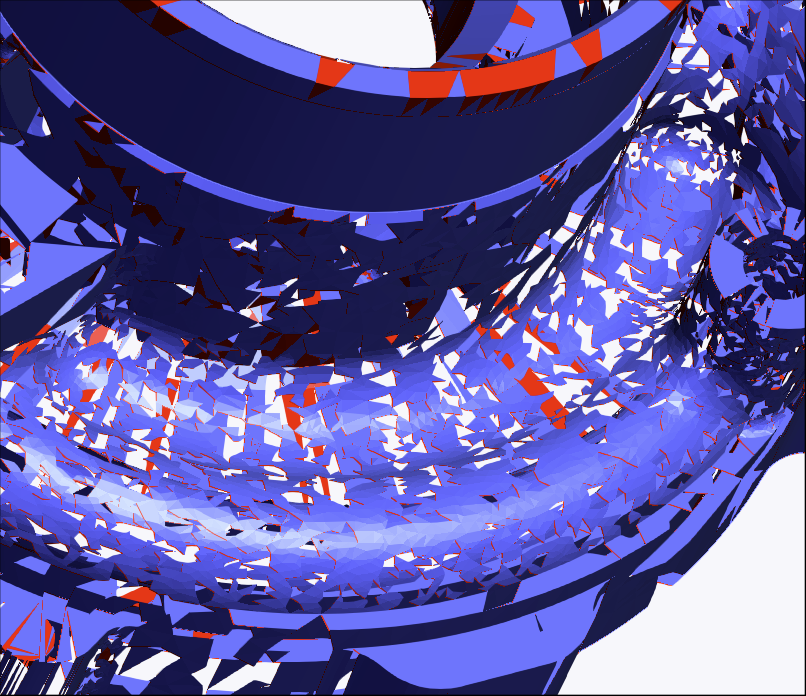}
    \caption{Artifacts in an assembly of parts. Missing triangles in the mesh indicate mismatched local measurements between a subpart tessellated on its own and the assembly tessellated as a whole.}
    \label{fig:file_format_2}
\end{figure}

\newpage
However, the strategy of converting models to a canonical file format (i.e., STL) has several drawbacks:
\begin{enumerate}
 \setlength\itemsep{0.5em}
    \item \textbf{Information Loss}: file format conversion often results in a loss of information  that can eliminate key details \citep{McHenry2009,McHenry2008}. This is particularly true of shapes defined by surfaces or curves that are subsequently converted into discrete (e.g., mesh, voxel) representations.
    \item\textbf{ Software Evolution}: as with any software artifact, conversion libraries evolve over time. Different versions of a library may utilize different meshing algorithms or different levels of numerical precision. A search engine's entire dataset of models may need to be regenerated when a new library version is released.
    \item \textbf{Parameter Dependency}:  libraries usually provide the user with a set of conversion parameters. Figure \ref{fig:tesselation_diff} shows the tessellation of a CAD model according to two slightly different parameter values. The resulting meshes have few triangles in common, creating problems for algorithms that utilize local neighborhood measurements that are sensitive to variations in tessellation. As discussed in later sections, some parameter settings can result in meshes containing degenerate  triangles that are  likely to be afflicted by numerical instability. 
    
    \item \textbf{Assemblies and Sub-parts}: tessellating an assembly can give different results than when the individual parts are separated and tessellated on their own.  Figure \ref{fig:file_format_2} shows matching between an assembly and one of its subparts, with the holes in the mesh showing triangles that fail to have similar local measurements. 
    
    \item \textbf{Post-conversion Matching}: as mentioned above, many models from OEM sites are available in several different file formats. Furthermore, users may convert a file and post it to a different location. Downloading multiple versions of the `same' model can lead to potential duplication in the database layer if the converted versions are not identical. 
\end{enumerate}

As a result of numerical issues (to be discussed in more detail in later sections), Physna is likely to move away from using STL as the canonical file format. These drawbacks of the STL file format are a problem for EXS and PIP search, which are the most important search types for Physna's enterprise customers. View-based methods (e.g., using image-based rendering and neural networks) are more robust with respect to the issues described above.


\subsection{Versioning}

3D search engines must deal with a variety of issues related to versioning and provenance. First, 3D models are often in a state of evolution. For example, authors frequently alter published models in order to fix errors, simplify geometry, or to introduce new features. In addition, websites occasionally change \textit{uniform resource locators} (``URLs'') or filenames. Both of these situations require the use of specialized detection routines and automated workflows in order to maintain a semblance of freshness.

Another type of versioning issue is an effect of model composition. For instance, many CAD artifacts are constructed through the use of one or more common (e.g., generic or reusable) sub-parts. As a result of composition, a given 3D artifact (e.g., CAD model of a pillow block bearing) may appear as a sub-part in otherwise unrelated assemblies published by different authors. It is not sufficient for a database to maintain a single source for a given model; rather, the mapping between models to sources must be many-to-many.

\begin{figure}[hb]
    \centering
    \includegraphics[width=4in]{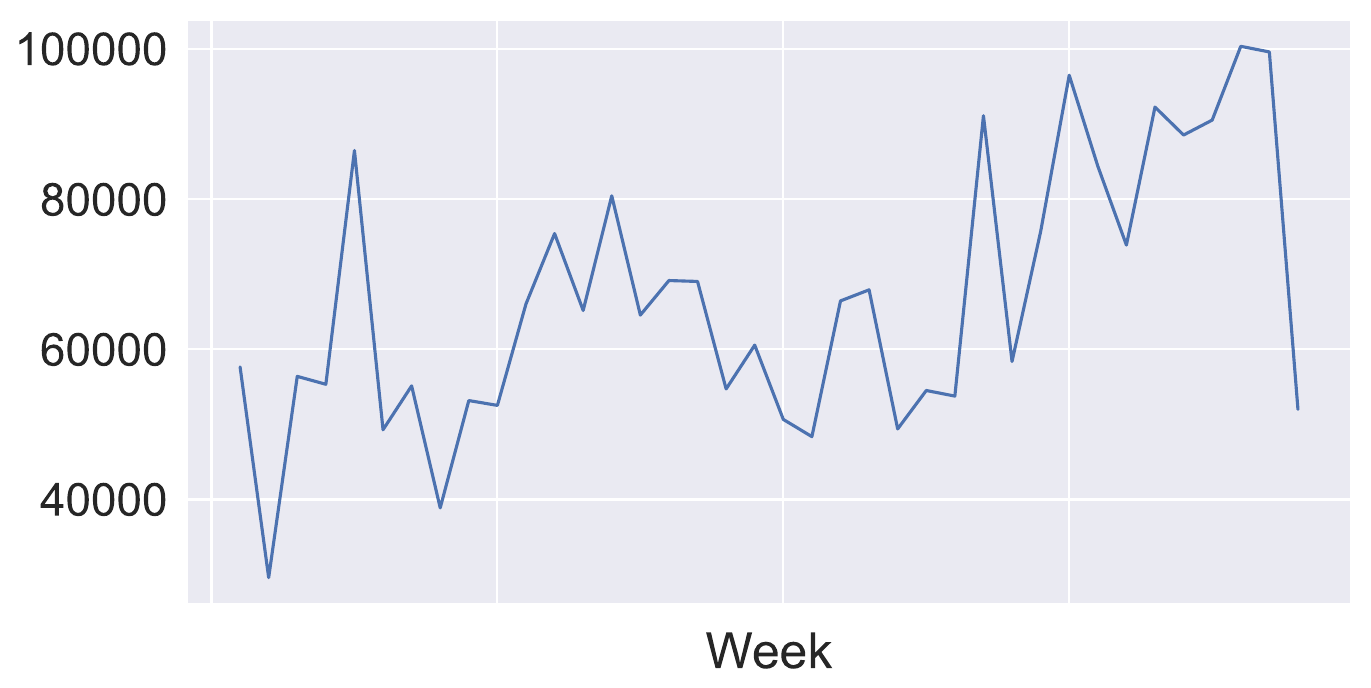}
    \caption{New models crawled via http request by week in 2022.}
    \label{fig:downloads}
\end{figure}

The Thangs search engine maintains version and provenance information for every 3D model in its various databases: the file format, file converter version number, ingestion history, and data source are all tracked explicitly in the database layer. The Thangs webcrawler also provides metadata such as acquisition/ingest timestamps. If a given 3D model is available on different websites (or through different file formats), the appropriate records are merged in the database to minimize duplication. Either md5 hashing or geometric similarity is used to match files. Figure \ref{fig:downloads} shows the number of unique (de-duplicated) model downloads per week from Jan to June of 2022.



\subsection{Restrictions on Crawling}
\label{sec:crawling_regulation}

There are significant legal risks involved in acquiring 3D models by crawling the world-wide web.  In the United States, web crawling is subject to a patchwork of laws \citep{Yannella2021,Ballon2020}, including: (1) contract; (2) tort (e.g., trespass to chattels, tortious interference); (3) unfair competition; (4) privacy; (5) intellectual property (e.g., copyright, trade secret), including the \textit{Digital Millennium Copyright Act} (``DMCA''); (6) the \textit{Computer Fraud and Abuse Act} (``CFAA''); (7) state anti-hacking statutes; (8) common law rights of publicity, and; (9) the \textit{Cybersecurity Information Sharing Act} (``CISA''). Although the majority of these laws expose web crawlers to civil liability, some of them provide for criminal remedies as well.

Many internet-accessible 3D models are found on  websites that restrict access through ``\textit{terms of use}'' (``TOU'') documents, ``robots.txt'' files, or registration requirements. Some courts have held that web scrapers can be bound by a website's TOU despite the lack of explicit acceptance (e.g., \citep{Register2004}). Accessing websites without permission may result in civil or criminal liability. For instance, Section 1030(a)(3) of the CFAA makes it 
a crime to intentionally access a computer either without authorization or beyond the scope of previously granted authority. Two recent cases are apposite:
\begin{enumerate}
 \setlength\itemsep{0.5em}
    \item In \textit{Van Buren v. United States} \citep{VanBuren2021} the Supreme Court provided guidance on the phrase ``exceeding authorized access,'' indicating that authorized users who abuse their privileges do not generally incur criminal liability.
    \item In \textit{HiQ v. LinkedIn}  \citep{HiQ2022} the 9th Circuit construed ``without authorization,'' concluding that sufficiently public data can be scraped without attracting liability under the CFAA. 
\end{enumerate}
These decisions are, of course, only concerned with one of the many causes of action that can be applied to web scraping. A search engine that does not incur liability under the CFAA may nevertheless be subject to civil liability on other grounds. For instance, if a web scraper violates provisions of a TOU, the website owner can invoke the relevant dispute resolution procedures, such as arbitration or litigation. Website owners whose servers have been impaired by web scrapers (e.g., degrading their performance or causing service outages) can seek redress through the tort of trespass to chattels  \citep{Christensen2020,Yannella2021}. In \textit{eBay Inc., v. Bidder's Edge, Inc.} \citep{Ebay2000}, a District Court found that an injunction was an appropriate remedy against a web crawler that impacted a website's servers (ignoring the site's \textit{robots.txt} file). The prospects of success in such an action vary between jurisdictions (see, for example, \citep{Register2004,Intel2003}).

Search engines should minimize their legal exposure by introducing an appropriate set of controls. The safest option in dealing with proprietary assets is to obtain consent from the owner. Some companies are happy to allow scraping in exchange for increased traffic, while others will not give permission under any circumstances. However, securing consent is difficult for startups as there are hundreds of OEMs and each agreement takes time to negotiate. 

\newpage

The table below summarizes the number of unique models that the Thangs crawlers have found on public and proprietary sites at the time of writing. 
\begin{itemize}
 \setlength\itemsep{0.5em}

    \item Around 5.9 million models are immediately accessible to web crawling, while others are accessible after obtaining permission or registering with an OEM site that allows downloads in its TOU.
    
    \item Many companies (e.g., Vention) will not grant access. 
    
    \item In addition, a few large repositories (e.g., GrabCAD) have not been assessed and are classified as ``\textit{to be done}'' (``TBD'').
\end{itemize}  These results indicate that a 3D search engine cannot rely on public websites to achieve significant coverage---in particular, gaining access to the Internet's vast store of engineering-oriented 3D artifacts will require consent from OEMs.
\begin{table}[ht]
\begin{center}
\begin{tabular}{@{}ll@{}}
\toprule
Access Specifier & Model Count \\
\midrule
Public & $5,882,528$  \\
Registration required & $3,995,397$ \\
Agreement required & $6,126,288$\\
Access prohibited & $4,232,093$\\
TBD & $7,000,000 +$\\
\hline
\end{tabular}
\label{table-sites}%
\end{center}
\end{table}

3D search engine operators should develop significant in-house expertise on web crawling not only to mitigate legal risk but also as a component of a comprehensive \textit{information security} (``IS'') program.  The Thangs search engine has been the target of web crawlers ranging from generic search engines (e.g., Google) to competing entities. In concert with comprehensive legal analysis, administrative and technical controls (e.g., domain blocking, rate limiting, monitoring of ``robots.txt'' files) should be introduced. Outside counsel may be used as a sanity check, but an evolving set of laws and target domains requires constant iteration. In some cases it may not be possible to block third parties from scraping public listings (see \citep{HiQ2022}), but their impact on computing resources can be minimized.

\newpage
\section{Intellectual Property}
\label{sec:intellectual_property}

\textit{Intellectual property} (``IP'') law can have an impact on the ability of search engines to acquire and retain data. This paper discusses three forms of IP: copyright, trademark, and trade secret.

\subsection{Copyright}

Copyright law grants authors the ability to control the dissemination and use of their works for a limited time. Various requirements limit copyright's scope (e.g., the works must be original and fixed in a tangible medium). 3D models will generally meet these requirements apart from rare edge cases, including: (1) models that were created in the absence of human participation, and;\footnote{See, for instance, the US Patent and Trademark Office's comments in  \citep{USPTO2014}.} (2) models that are the result of a process (e.g., 3D scanning) that lacks a  minimum amount of creativity (see, for example, \citep{Salkeld1975,Dasari2013}). Copyright for computer generated works has been a subject of debate for decades (e.g., \citep{Samuelson1986}). 

The copyright law of the United States includes both civil penalties (e.g., 17 U.S.C. \textsection 501-505) and criminal penalties (17 U.S.C. \textsection 506). Civil remedies include forfeiture of profits, restitution of losses, statutory damages, attorney's fees, and confiscation/destruction of infringing items (i.e., a ``writ of seizure''). Plaintiffs can also obtain either a temporary or permanent injunction (17 U.S.C. \textsection 502a) that prohibits further infringement. 

Apart from the general criminal provisions in Section $506$, search engine operators should also be aware of 17 U.S.C. \textsection 1201, which prohibits them from circumventing a ``technological measure that effectively controls access'' to a protected work. In addition to imposing restrictions on web crawling activities, this provision essentially prohibits 3D search engines from displaying ``direct download'' hyperlinks that bypass registration/login requirements or other technical controls.

Assessing a search engine's exposure to risk under copyright requires mapping out its \textit{data flows}. In general, a 3D search engine: 
\begin{enumerate}
 \setlength\itemsep{0.5em}
    \item acquires 3D data and related metadata (e.g., creation date, version, written description, keywords);
    
    \item (optionally) performs file format conversion to create secondary copies of a model in a canonical file format;
    
    \item transforms the 3D data into a set of intermediate representations that support some form of search, such as mesh fingerprints \citep{Warner2017}, 2D images, or locality-sensitive hash signatures  \citep{Lee2012};
    
    \item extracts statistical information about the data (e.g., triangle count, bounding volumes), and;
    
    \item displays representations of the original and/or transformed data to its users (e.g., preview images, 3D-enabled thumbnails).
\end{enumerate} 
\newpage

As a result, a 3D search engine typically maintains a database that includes copies of original works, transformed works, statistical information, and metadata. Legal counsel should review the entire dataflow, paying particular attention to issues of jurisdiction (e.g.,  material obtained from a company that can rely upon the European Union's Database Directive). 

Under US law, there are two main copyright issues that arise when multiple works are scraped from an internet-accessible collection \citep{Ballon2020}: (1) violations of copyright in the individual works, and; (2) violations of copyright in the entire collection. Copying a substantial amount of a collection can open up a web scraper to liability on both fronts.  While it would be preferable to avoid storing original works, access to those artifacts is often required for debugging. However, textual descriptions accompanying a 3D model are usually not protected (e.g., technical specifications that describe basic characteristics of an artifact).

Operators of 3D search engines can invoke several defences to copyright infringement claims, including ``fair use'' (17 U.S.C. \textsection 107). In \textit{Authors Guild v. Google Inc.} \citep{Authors2015}, the Court of Appeals for the 2nd Circuit examined copyright claims against Google Books. While the court agreed that Google had infringed copyright by engaging in the unauthorized digitization of books, the fact that the copying was ``transformative'' (i.e., done for the purpose of creating a search engine) made it a non-infringing fair use; nor was this result negated by the fact that Google was a commercial entity with a profit motive (see also \citep{Kelly2002,Perfect2007,Stokes2019}). 

While \textit{Authors Guild} is useful as precedent, it is worth noting that fair use is determined on a case-by-case basis in an open-ended and context-sensitive inquiry  \citep{Leaffer2019}. Operators of 3D search engines should retain experienced legal counsel to advise on fair use, improper appropriation, intermediate copying, and other issues. A ``notice and take-down'' policy must be imposed, and the search engine software must be designed so that content can be taken offline easily. Physna's back-end software teams have created a deletion pipeline that, given a model identifier, automates the takedown process. While it is tempting to only store metadata or derived data, experience shows that having the original 3D model files is essential for debugging and for investigating inquiries or complaints from users. Data for a given 3D model will generally exist at various places within the search engine's IT infrastructure, including training datasets, bug reports, and log files.

There are additional liability shields available for both internet-scale search engines and content hosting platforms (e.g., Thingiverse). Organizations that operate such services may be able to avail themselves of the ``safe harbor'' provisions of the DMCA (17 U.S.C. \textsection 512) and  Section 230 of the \textit{Communications Decency Act} (``CDA''). Under the DMCA, \textit{online service providers} (``OSPs'') who meet certain criteria are granted significant exemptions from liability arising from copyright infringement. For instance, Google successfully invoked one of the four statutory safe harbors to defend against infringement claims related to copyrighted works in its cache \citep{Field2006}. The CDA allows certain entities to avoid causes of action under state IP laws. 

\subsection{Trademark}

3D search engine operators must also consider trademark issues, particularly if corporate logos or trademarked terms are scraped. Trademark infringement analysis is a specialized area of law with a large body of cases (e.g., \citep{WhenU2005}). In general, it is not enough for a plaintiff to show that their mark was used without authorization ``in commerce''---rather, they must also show that the use was likely to cause confusion as to affiliation or origin, sponsorship, or approval. Although the case law is relatively sparse, the Court of Appeals for the Second Circuit considered trademark use in the context of search engines in \textit{RescueCom v. Google} \citep{Rescuecom2009}. A search engine can generally avoid liability under trademark law by avoiding uses of trademarks in a manner that has the potential to deceive or confuse consumers.

\subsection{Trade Secret}

Trade secret is typically not an issue if a search engine crawls pages that are explicitly made available to the public. However, there are at least two edge cases to be aware of: (1) where taking the entire content of a website's database of 3D models puts the scraper in a position to obtain trade secrets, and; (2) where the scraper bypasses access controls and obtains trade secret information. Both of these scenarios can be avoided with proper due diligence.



\section{Trustworthy Computing}

For a search engine to be successful it is not enough for it to provide value through its basic functionality. It must also gain the trust of its users. This section describes four issues faced by 3D search engines: privacy, security, inappropriate content, and remixing of digital assets.

%
\subsection{Privacy}

Content hosting sites like Thangs or Thingiverse allow users to register and upload their own models. This design choice exposes the platform to data protection statutes such as the European Union's \textit{General Data Protection Regulation} (``GDPR''). The platform operator must introduce safeguards to demonstrate compliance with these sources of law, including restrictions on collection, use, disclosure, and retention of \textit{personal information} (``PI''). 

Many of the privacy issues faced by these sites are generic and easily handled by existing practices. However, there are some subtle privacy risks that are unique to the domain. For instance, users who upload models can embed PI into the 3D model data and/or metadata (e.g., the user could place PI into a `description' field, or they could use a 3D extrusion utility to convert text containing PI into a 3D model). 

A 3D search engine is not obligated to offer content hosting functionality. However, the issue of personal information within 3D model data is still a concern. Third parties can embed PI into 3D models and seed them on websites that are subsequently scraped. As soon as PI resides within the search engine's database, the search engine operator has to worry about various data protection laws.

%
\subsection{Security}

3D search engines face all of the typical security risks of an internet-accessible software application \citep{VanOorschot2020,WalterWilliams2022}, including denial of service and SQL injection attacks. 3D content hosting platforms have an even larger threat profile, due to the functionality offered to users. This is not a theoretical issue, as the Thangs search engine has been faced with malicious users on several occasions. These experiences have prompted improved security controls, including automated monitoring routines that attempt to detect when users engage in suspicious behavior. Rate limiting and IP blocking are automatically invoked when the search engine encounters unusual access patterns.

Multimedia search engines also face unique security threats that are not discussed in the general security literature. As described in  Section \ref{sec:background_attacks}, a significant amount of effort has been devoted to security threats in the context of \textit{additive manufacturing} (``AM''). Previous works have already analyzed  attacks that make use of the STL file format (e.g., \citep{Sturm2017}). For instance, an attacker can modify an STL to make it unsuitable for 3D printing through by creating internal voids.


Although the existing research on security in AM systems offers useful insights, it is not a reliable guide to the security threats facing 3D search engines or 3D content hosting platforms. While both types of information system are vulnerable to attacks by internal actors (e.g., employees), they differ with respect to external threats. 3D content hosting platforms (e.g., Thingiverse) are vulnerable to direct upload attacks. A common pattern is for an attacker to: (1) download a 3D model; (2) alter it to create several modified versions, and; (3) upload the modified versions. The Thangs security team has been forced to deal with this type of attack on several occasions.

In contrast, external actors do not have direct upload access to a 3D search engine. They are relegated to uploading content indirectly by seeding models on a site that is subsequently crawled. Actors with knowledge of a particular shape similarity algorithm, for instance, can seed data that degrades search performance or search quality. Analysis of operational data from Thangs demonstrates that few users will view more than the top 15 results to a query. An attacker who manages to insert high-ranking false positives (e.g., models that are closely related to an existing model but modified in a malicious manner) can easily surface these models to the user if they are included in the set of top-scoring results.

%
\subsection{Inappropriate Content}

A 3D web crawler will routinely encounter material that is considered inappropriate by its user community (e.g., obscene content). The same is true of 3D content hosting platforms, which provide a convenient means of uploading textual or 3D data that does not meet community standards. Uploads of inappropriate content can occur for several reasons, including: (1) differing assessments of what material is appropriate; (2) failure to read the TOU, or; (3) malicious intent. As an empirical matter, inappropriate content is routinely encountered by Thangs either through direct uploads or as a result of web crawling. Search engine operators must be aware that this type of content is a significant source of risk to public reputation (i.e., goodwill).

\begin{figure}[ht]
    \centering
    \includegraphics[width=4.4in]{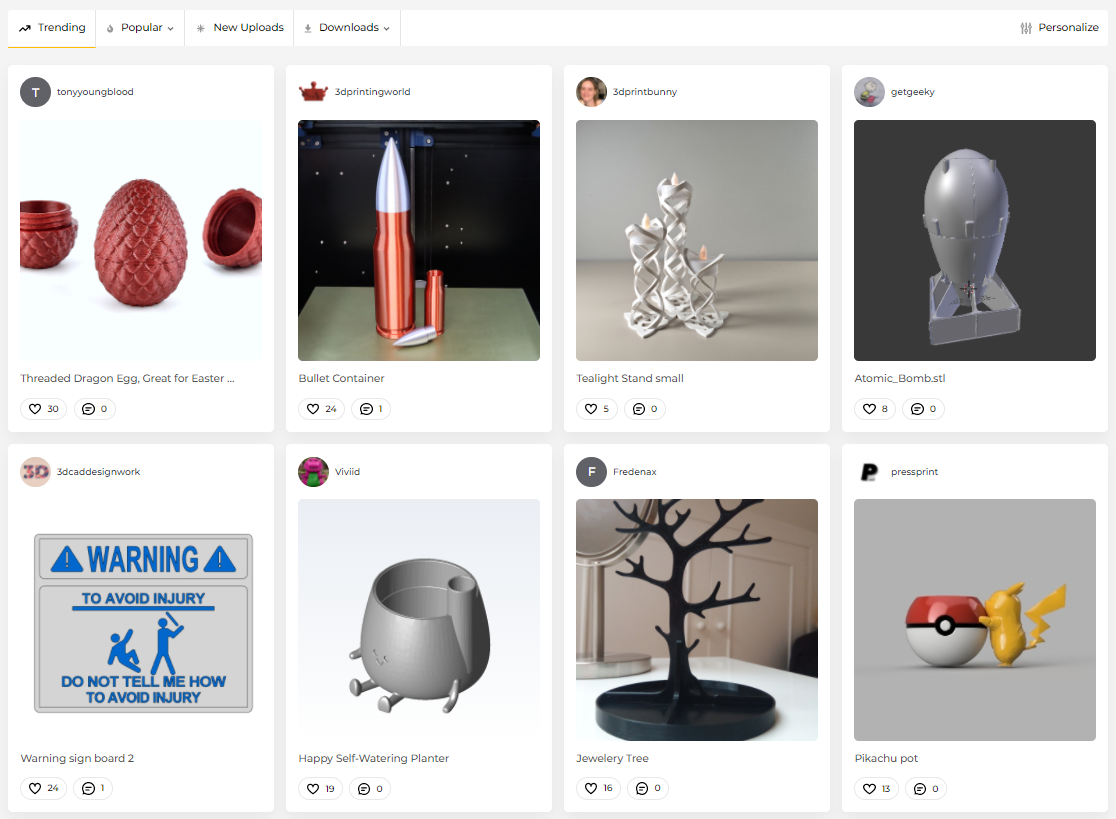}
    \caption{Trending Models on the landing page of the Thangs Search Engine. Inappropriate content should not appear on this page.}
    \label{fig:trending}
\end{figure}

\newpage
Detection of inappropriate text is much easier than detection of inappropriate 3D content. Thangs uses Google NLP and Vision to prevent inappropriate models from being shown on the home page; users can also provide feedback through the ``report model'' button. Thangs allows content that is \textit{not safe for work} (``NSFW''), but it is not shown in trending results (see Figure \ref{fig:trending}). Inappropriate search terms are blocked (blacklisted) so that they cannot be executed. Inappropriate models are a small percentage of the overall database ($\ll1\%$), but they have an impact on user trust.

\subsection{Copying and Remixing}

One major set of issues facing 3D content hosting platforms arises from the copying and remixing of user-generated content. As mentioned in Section \ref{sec:background_remixing}, the existing research (e.g., \citep{Papadimitriou2015,Flath2017}) shows that remixing design strategies are used in a large percentage of 3D artifacts on Thingiverse. \citet{Flath2017} describe how Thingiverse provided explicit support for remixing through provide TBS mechanisms like content tagging, categorization, as well as a unique `customization' workflow. Since a large number of users want the ability to find and remix 3D content, content creation platforms should provide support for remixing as an important use case.

However, a 3D search engine or content hosting platform that indexes both open source and proprietary content has to consider the needs of users who do not wish their content to be copied or remixed. In the most obvious scenario, OEMs will contact a search engine operator to remove links to unauthorized copies of their models on third party sites. The same ``notice and takedown'' workflow will be invoked if the OEM finds `remixed' models that use a substantial amount of their content. 

Both Thangs and Thingiverse allow users to define their own license terms when they upload models. As a result, the platforms must provide support for remixing, but also for users who wish to enforce their legal rights. This is not a merely theoretical issue---rather, copying and remixing of 3D content is a major challenge for the Thangs search engine in practice. The most common scenario is as follows: 
\begin{enumerate}
  \setlength\itemsep{0.5em}
    \item user $A$ uploads model $M$ directly to the platform;
    \item user $B$ downloads $M$ and alters it in some simple manner to produce a derivative work $M'$;
    \item user $B$ uploads $M'$ to the platform under their account, and;
    \item user $A$ eventually discovers $M'$ and posts negative comments about the platform on social media.
\end{enumerate}

This type of scenario can also impact a search engine if user $B$ seeds model $M'$ on a site that is subsequently crawled.

\begin{figure}[hb]
     \centering
     \includegraphics[width=6.2in]{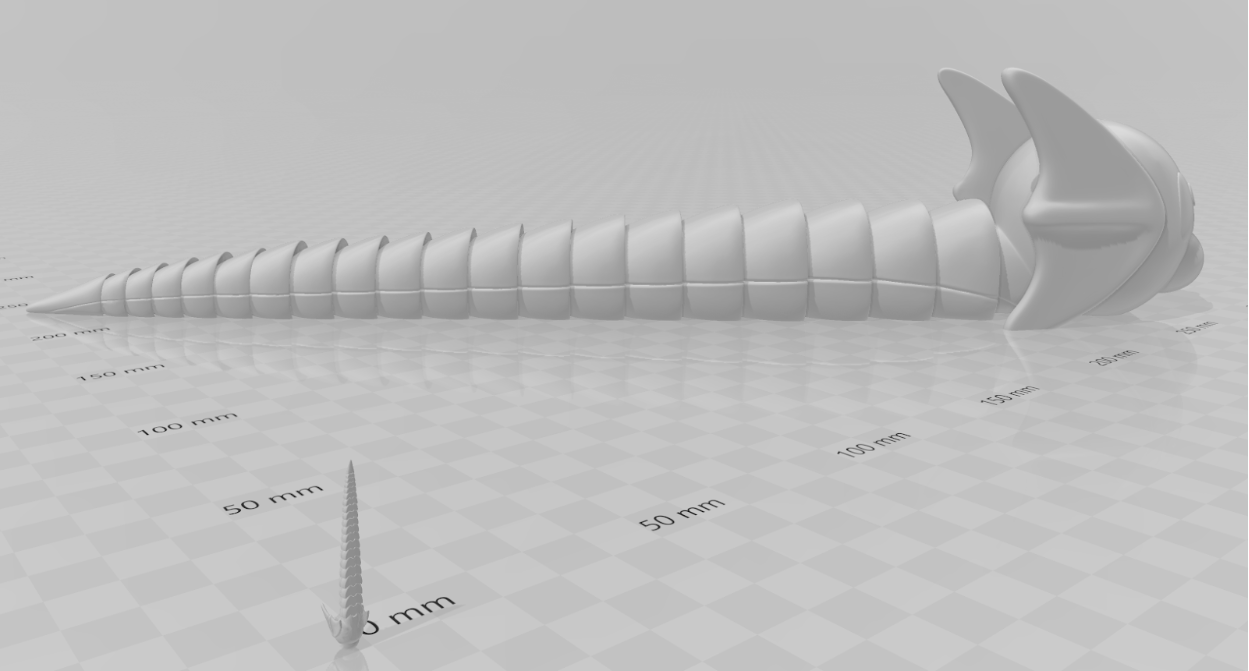}
     \caption{A ``real world'' example of a copied work that was re-uploaded. The smaller model was created by applying transformation, rotation, and scaling to the larger model.}
     \label{fig:dratini}
 \end{figure}

 In some cases malicious users will even create trivial derivative works by applying affine transformations to the original 3D model (e.g., uniform scaling, translation). Figure \ref{fig:dratini} shows a real-world example of this type of content duplication. An original work by user $A$ was downloaded  by user $B$, who created a trivial derivative work through the application of affine (i.e., translation, rotation, and scaling) transformations. User $A$ discovered the modified version through the ``view related models'' feature and contacted Physna to complain. This particular instance was easily detected through view-based similarity search.

\newpage

\begin{figure}[hb]
    \centering
    \includegraphics[width=6in]{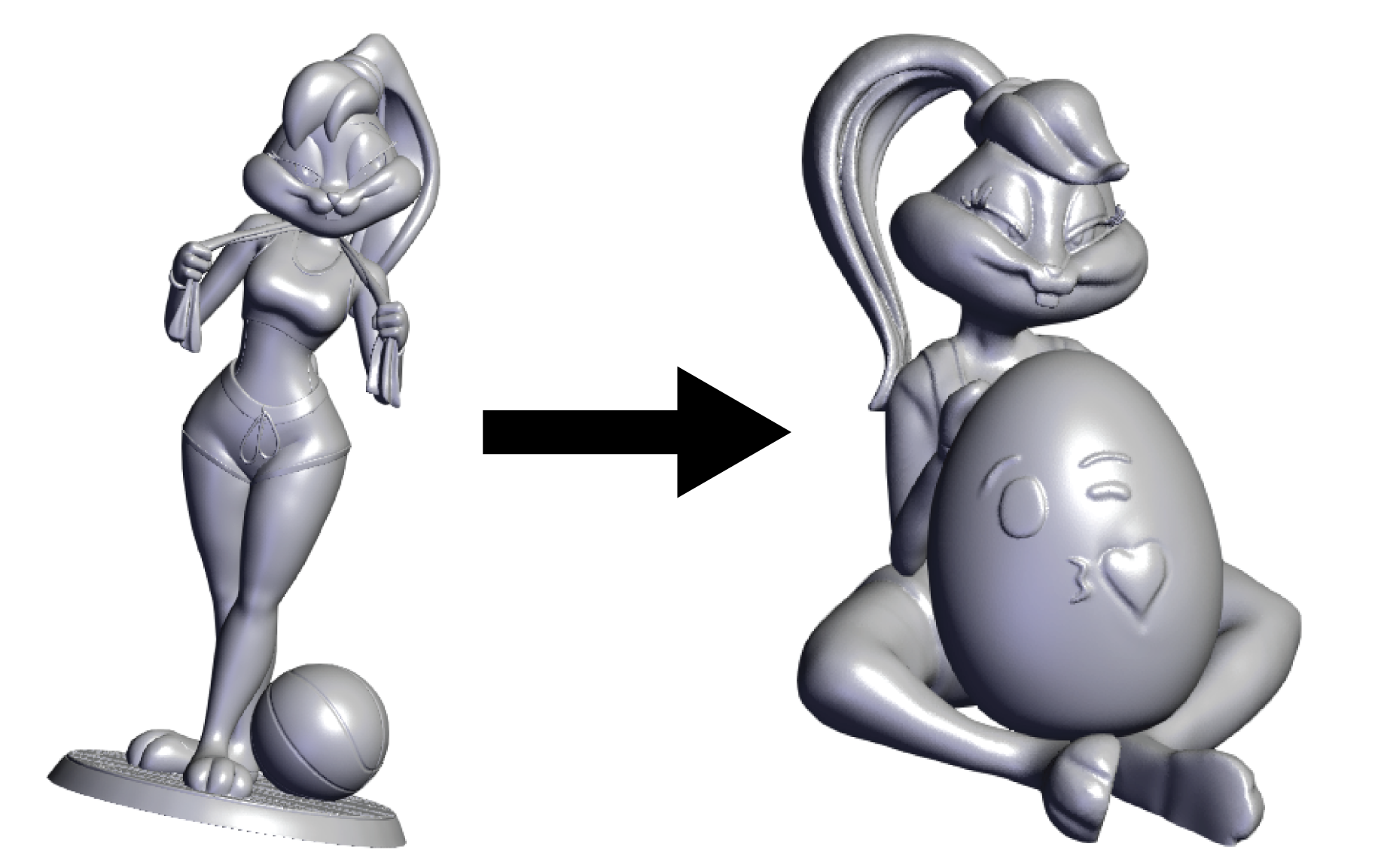}
    \caption{A novel work by a Thangs user (left) was downloaded and altered to create a derivative work (right) that was subsequently uploaded to Thangs without crediting the original author. Detecting this sort of `remix' is a challenging problem.}
    \label{fig:lola}
\end{figure}

 Figure \ref{fig:lola} shows a more sophisticated ``real world'' example of model remixing. The model on the left is a novel work by a Thangs user, while the model on the right is a derivative work that uses a substantial portion of the original geometry. When the original author discovered the remixed model, she issued a series of indignant social media posts about the ability of the platform to detect remixes.

In summary, copying and remixing can be a significant problem for a search engine or content hosting site. Many users expect a platform to identify derivative works and bring them to the attention of the original author. In the case of Thangs, some modifications are caught by Physna's PIP search techniques, while others can be detected through 3DSS. However, there are hard cases that are not caught by any of Physna's existing search methods. Detecting remixed works by geometry is a problem that requires further research and development by the 3D IR community.



\newpage
\section{Information Retrieval}

This section covers some important challenges for large-scale information retrieval of 3D models, including data integrity errors, common/generic geometry, and collisions in local measurements.

\subsection{Search Variants}

\begin{figure}[hb]
    \centering
    \includegraphics[width=3.6in]{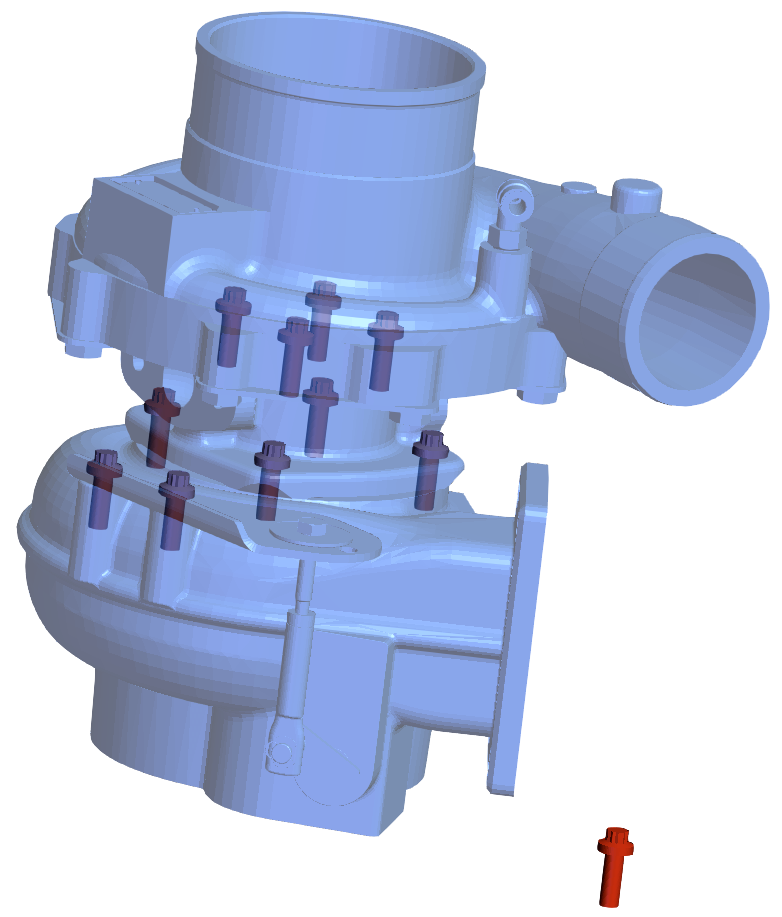}
    \caption{An example of part-in-part (``PIP'') search for a bolt in an assembly. Assemblies may contain a given part multiple times in different locations.}
    \label{fig:part-in-part}
\end{figure}

As mentioned above, a 3D search engine can support one of several search modalities: 
\begin{enumerate}
  \setlength\itemsep{0.5em}

    \item  \textit{text-based search} (``TBS''), which allows users to find relevant models by entering metadata (e.g., search terms like `robot');
    
    \item \textit{exact search} (``EXS''), which allows users to find a specific model in the database by uploading a query model;
    
    \item  \textit{similarity search} (``3DSS''), which returns a set of models that are similar to the query model, and;
    
    \item \textit{part-in-part} (``PIP'') search \citep{Muraleedharan2019}, which allows a user to upload a query model and search for models that contain it as a component (see Figure \ref{fig:part-in-part}). 
\end{enumerate}

With reference to Thangs, TBS is straightforward and by far the most popular with typical users at over 90\% of public search volume.
Since Thangs is both a model hosting site and search engine, search results are bifurcated into `internal' (hosted) and `external' (crawled) entries. Filters can be invoked to find models that: (1) are watertight; (2) have consistent normals; (3) are of a particular file format, or; (4) come from a particular internet domain. Usage data indicates that these filters are extremely important to users, with the first two filter options being the most popular.

3DSS is currently implemented using a view-based, multi-stage filtering approach that uses a combination of techniques from image-based rendering, deep neural networks, and computational geometry. Each model in the Thangs dataset is pre-processed using a surface sampling method that produces a tensor of fixed dimensions. An embedding vector of floating point values is created for each model by feeding the input tensor into a deep neural network. Given a database of such vectors, a 3DSS query is handled by: (1) processing the query object to obtain its embedding vector; (2) searching a database of embedding vectors for the nearest neighbors, and; (3) applying additional filters to the candidate models produced by the first two steps.

\begin{figure}[ht!]
    \centering
    \includegraphics[width=5.6in]{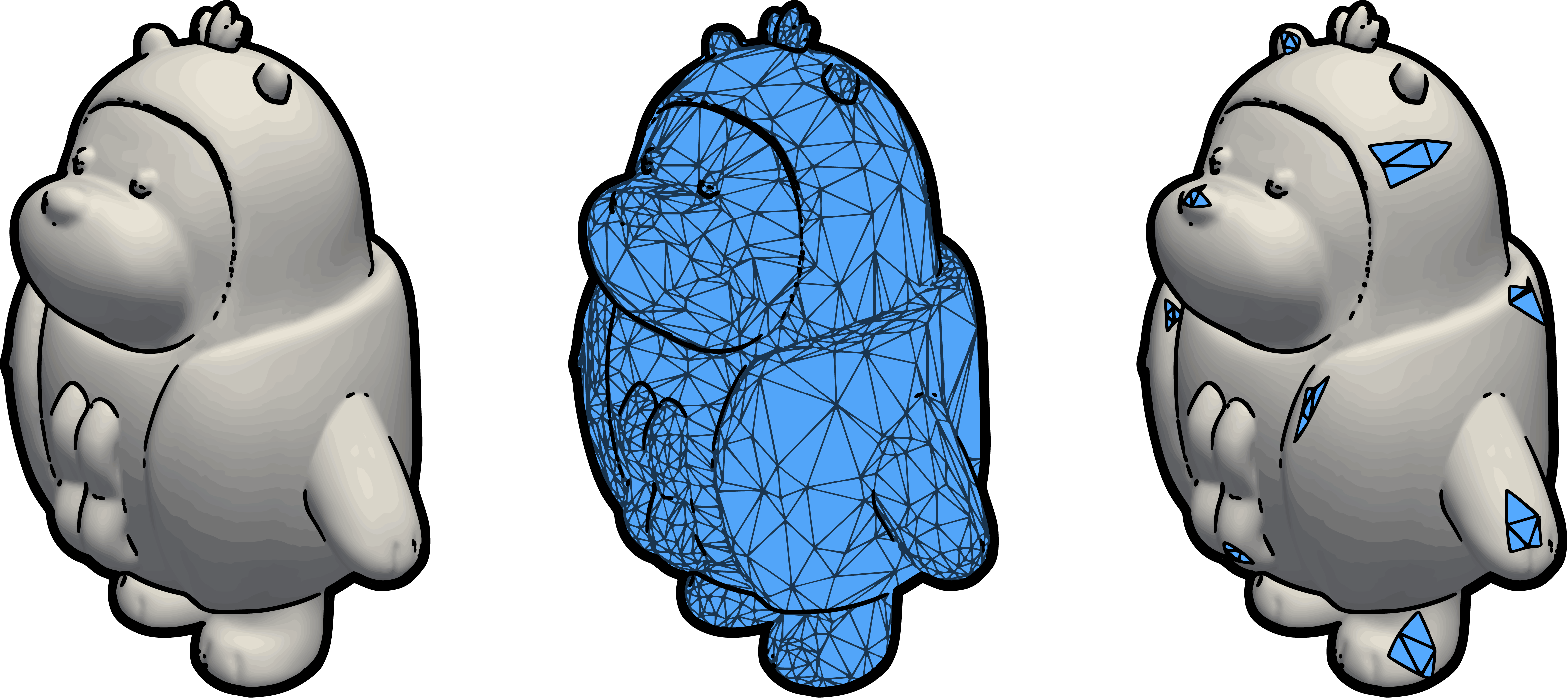}
    \caption{The Thangs search engine decomposes a model into a multiset of geometric ``words.'' (Middle) A relatively small number of words describe the global shape. (Right) A large number of primitive shape descriptors describe the shape locally.}
    \label{fig:thangs2}
\end{figure}

PIP and EXS are implemented with a multi-stage ``\textit{bag of words}'' (``BOW'') approach \citep{Warner2017} that uses shape descriptors (e.g., \citep{Corney2002}) in combination with a local measurement for each triangle in a mesh (Figure \ref{fig:thangs2}). A model is represented (i.e., fingerprinted) by a multiset (bag) of local measurements. A basic measure of similarity between two models is obtained by computing the overlap between their multisets; however, this type of similarity measurement is not sufficient for search in large collections. Thangs uses two sets of additional descriptors: (1) mid-level descriptors that combine information from the mesh with local measurement data, and; (2) global shape descriptors  that do not make use of local measurement data at all. A sequence of filtering stages compare model descriptors in order to winnow down the set of candidate objects to a concise list of search results that can be surfaced to the user.

 In contrast to ``\textit{bag of features}'' (``BOF'') approaches \citep{Csurka04visualcategorization,Liu06shapetopics}, the geometric words used in the Thangs BOW approach are not quantized by fitting them into a binning scheme derived from the distribution of words in the entire dataset (e.g., \citep{Ohbuchi2008,Furuya2015}).\footnote{Some authors use the term ``bag of words'' for what we describe as ``bag of features'' (e.g., \citep{Li2008}).} Quantization of this sort causes significant problems (e.g., false positives) when a dataset contains millions of models. Furthermore, CAD models and sculpted models have different data distributions, making it difficult to define a suitable binning scheme for a mixed dataset.

PIP is the most important search variant for Physna's enterprise customers and a distinguishing feature for Thangs. Because of its importance to the company, this section concentrates largely on data management issues facing PIP search.

\subsection{Bag-of-Words Approaches to Information Retrieval}
\label{sec:bow_retrieval}

There are many ways to define a notion of similarity for 3D shapes. They may be considered similar semantically if they represent similar conceptual subjects \citep{wang2015sketch}, stylistically if they bear visual resemblance \citep{lian2010visual}, mechanically if they bear coupled or mated parts \citep{Muraleedharan2019}, or abstractly if they contain significant isometric subsurfaces \citep{lian2010non}. 

One popular method of accommodating multiple context-dependent notions of similarity is the aforementioned \textit{bag-of-words} (``BOW'') approach, in which an object is mapped to a multiset (bag) of context-specific descriptors (words). In text-based search \citep{tsai2012bag} a text document is reduced to a multiset that contains its composite words. Texts can then be compared by simple operations (e.g., multiset Jaccard Similarity) or they can be used as input for more sophisticated models. For example, multisets can be considered as linear combinations and embedded in a vector space (as in word2vec \citep{word2vec}) for metric clustering algorithms or spatial tree search algorithms. Alternatively, word frequencies may be analysed to build classifiers (e.g., spam filtering) \citep{spamham}.



One of difficulties facing BOW methods is that  the same thing can be expressed in multiple ways; nearly identical sentiments may have many different texts that yield different representations as BOWs. Various techniques (e.g., word2vec) have been developed to map data into high dimensional vector spaces that preserve semantic relationships.

The BOW approach is applicable for any category of objects that are composed of discretizable sub-objects from some finite collection \citep{tsai2012bag, 5414415, memon20193d}. When applied to 3D shapes, the meaning of a shape comes from sub-shapes or geometric primitives (words) that are relatively positioned (syntax) into an entire model (text) \citep{5414415}.
Just as with text, alternative representations of the same object do not generate the same BOW unless some equivocation or notion of similarity is established between some of the primitives.

\newpage
\subsection{Bag-of-Words Advantages}

The BOW approach to shape retrieval has several  advantages, including: (1) flexibility of similarity notion; (2) respect for sub-object inclusion; (3) independence of word counts, which makes models flexibile to on-line changes, and; (4) natural parallelization. BOW approaches can accommodate any notion of similarity, provided one can compute the relevant geometric words from the shapes. 

Perhaps the most important property of the BOW framework is that it respects sub-object relationships. For example, if $S$ is a text which is a passage of some larger text $T$, then all of the words of $S$ are also in $T$ [i.e., bag$(S) \subseteq $ bag$(T)$]. This property is critical for PIP search, as it ensures that the bag for a sub-part is always contained in the bag for its super-part.

The Thangs search engine makes use of several types of geometric words, including: (1) \textit{local words} which are defined for small subshapes (e.g., \citep{Warner2017}), and; (2) \textit{global words} encompassing the entire shape (e.g., \citep{Corney2002}). As shown in Figure \ref{fig:thangs2}, each facet in a mesh is assigned a local word  derived from local neighborhood measurements. Physna constructs search variants by tailoring the geometric words to the relevant notion of similarity.

BOW models can be maintained with a great deal of online flexibility. While it is common to represent bags as vectors of word counts over a finite dimensional vector space with a fixed basis set of words \citep{1314502}, the bags can also be viewed as sparse vectors over a dynamically defined vector space in which particular basis words can be added or removed. Since the bags are compared as counts over a basis, any particular basis elements can be added or removed without disruption to the larger representation. Similarly, the independence of the geometric words makes the search retrieval parallelizable.

\subsection{Bag-of-Words Model Challenges}
\label{sec:storage}

Numerous difficulties arise when a BOW-based shape retrieval algorithm is applied to extremely large datasets like the Thangs database. This is a function of: (1) the volume of data accessed in response to a query, and; (2) limitations of geometric word descriptors.

\subsubsection{Storage Requirements}

\begin{figure}[ht]
    \centering
    \includegraphics[width=5in]{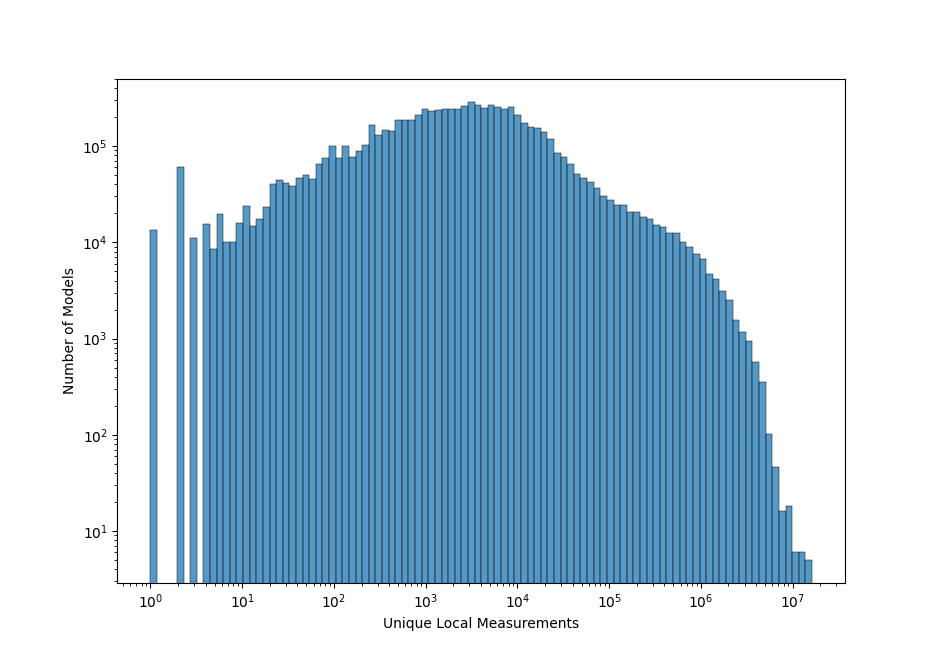}
    \caption{Distribution of local measurements per model on a dataset of 8 million models.}
    \label{fig:word_histogram}
\end{figure}

3D models vary greatly in both complexity and file size. Large meshes (e.g., those representing protein structures) may contain upwards of 40 million triangles in files that require gigabytes of storage. Given that the number of geometric words is usually a significant fraction of a model's triangle count (see Figure \ref{fig:word_histogram}), a BOW approach can be costly in terms of storage requirements. A database of 20 million models can easily require hundreds of terabytes for the original 3D model files; storing derived data (e.g., conversions to different file formats, mesh fingerprints, sets of 2D images) exacerbates these storage requirements, but having access to the full set of artifacts is required for debugging.

Large datasets are not an insuperable challenge for a search engine; rather, the issue is that BOW-based queries often require examination of a large percentage of total data. In the worst case (e.g., PIP search for small parts that have frequently occurring local measurements) they are analogous to querying for all the passages from a full set of encyclopedias.

\newpage
Only a fraction of a large dataset can be stored in \textit{random access memory} (``RAM''), making traditional in-memory data structures like trees less effective due to: (1) the latency involved in accessing distributed, persistent storage, and; (2) the impact of chasing pointers on the cache hierarchy. The latter issue is noticeable in multi-core/multi-threaded processing on the CPU. 

Furthermore, the size of a geometric word has implications for the full hardware/software stack. For instance, larger word representations may be a poor fit for hardware (e.g., DRAM) boundaries, and/or database indexing mechanisms. These issues are not noticeable with small datasets, but they become a major consideration when datasets run into the tens/hundreds of millions of models (i.e., petabyte scale).


\subsubsection{Data Integrity Errors}

Geometric words are subject to many sources of error that are not present in text-based approaches. Errors introduced by representation or file format conversion \citep{McHenry2009} can be preserved (or amplified) in the computation of geometric words. The latter can be afflicted by numerical noise or instability (particularly if geometric words are not continuous in shape parameters).

\begin{figure}[ht]
    \centering
    \includegraphics[width=6.2in]{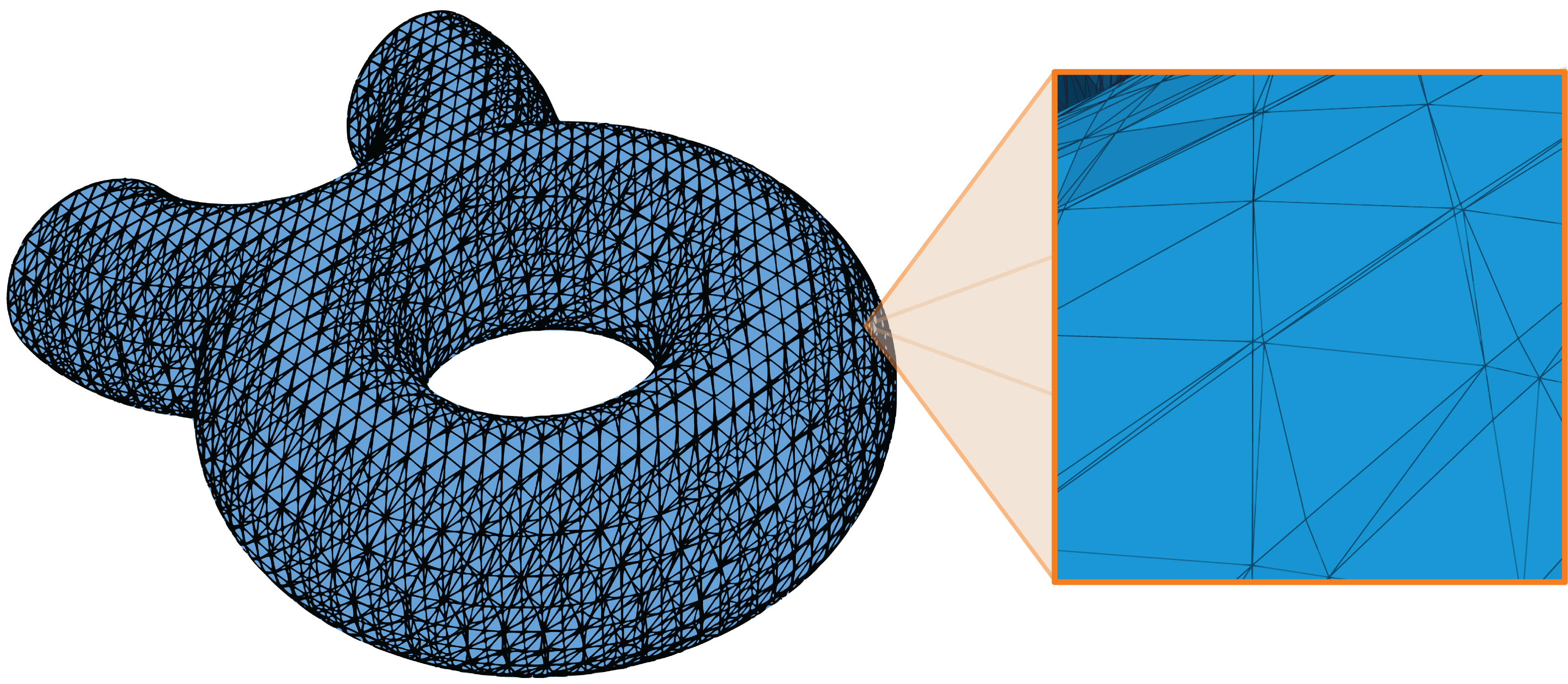}
    \caption{Degenerate triangles formed by tessellating a torus with a marching cubes algorithm.}
    \label{fig:nulls}
\end{figure}

Another source of error comes from malformed facets that are created when translating a model from one shape representation to another. For instance, marching-cubes approaches to tessellation will often yield extremely slim triangles whose local measurements (i.e., word representations) will be affected by numerical artifacts (Figure \ref{fig:nulls}). A 3D search engine will frequently encounter triangles whose local measurements (e.g., area, perimeter) have been rounded to $0$. This can cause problems for some shape similarity algorithms, and also for quality/sanity checks built into an ingestion pipeline.

\begin{figure}[ht]
    \centering
    \includegraphics[width=6.2in]{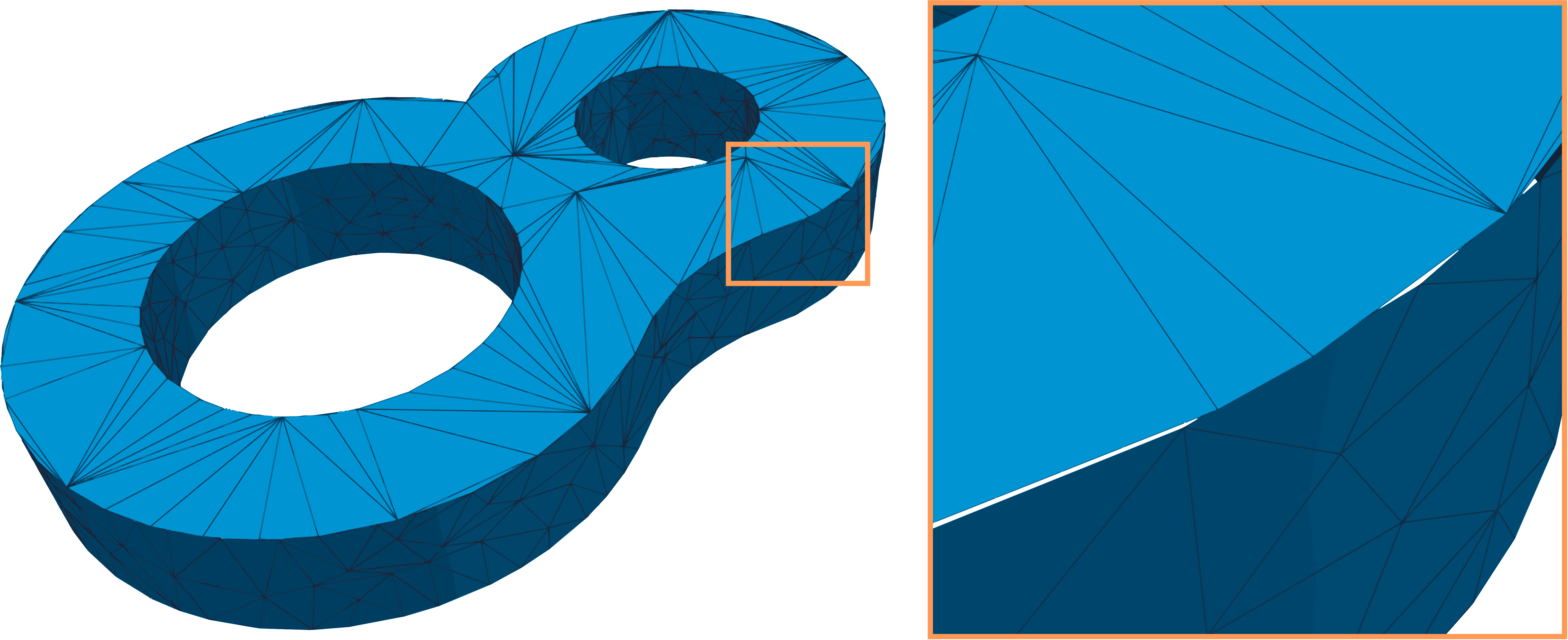}
    \caption{A mesh that contains unstitched triangles along a boundary.}
    \label{fig:unstiched_boundary}
\end{figure}

Many mesh-based 3D artifacts are subject to non-numerical (e.g., topological) errors in their mesh representation, whether as a result of software defects or poor modeling techniques. It is common to encounter basic errors like non-manifold edges, flipped normals, coplanar self-intersections, and duplicated facets or vertices (see \citep{Etigowni2021,Osada2001,DBLP:journals/corr/ZhouJ16}). Figure \ref{fig:unstiched_boundary} shows an example of a mesh in which triangle edges are not properly stitched.

It is not easy to automatically repair these types of issues (see \citep{Attene2013,Ju2009}) and human intervention is often required.  Even worse, the repairs may cause problems for search algorithms. For instance, re-tessellation can result in a different BOW than that provided by the original input mesh; manipulation of malformed triangles can affect words/measurements for neighboring triangles, potentially losing features relevant to PIP search.
 
Ameliorating data integrity errors is particularly difficult for a search engine, since alterations to local copies will introduces discrepancies with the corresponding source models on the internet. An 3D search engine is not tasked with quality control, but rather with indexing the models that exist ``in the wild.'' Given this mission, a 3D search engine should only perform alterations to a model when data integrity errors have the potential to negatively impact the information retrieval scheme itself (e.g., degrade indexing performance or cause preview image rendering errors). It may, however, surface warnings to users if a mesh has structural issues. In some cases it is best to reject entire files; for instance, Physna has found that extreme values in vertex positions are a signal that a 3D model file is likely damaged and ultimately not worth indexing.

\subsubsection{Common Geometric Structures}

\begin{figure}[ht]
    \centering
    \includegraphics[width=6in]{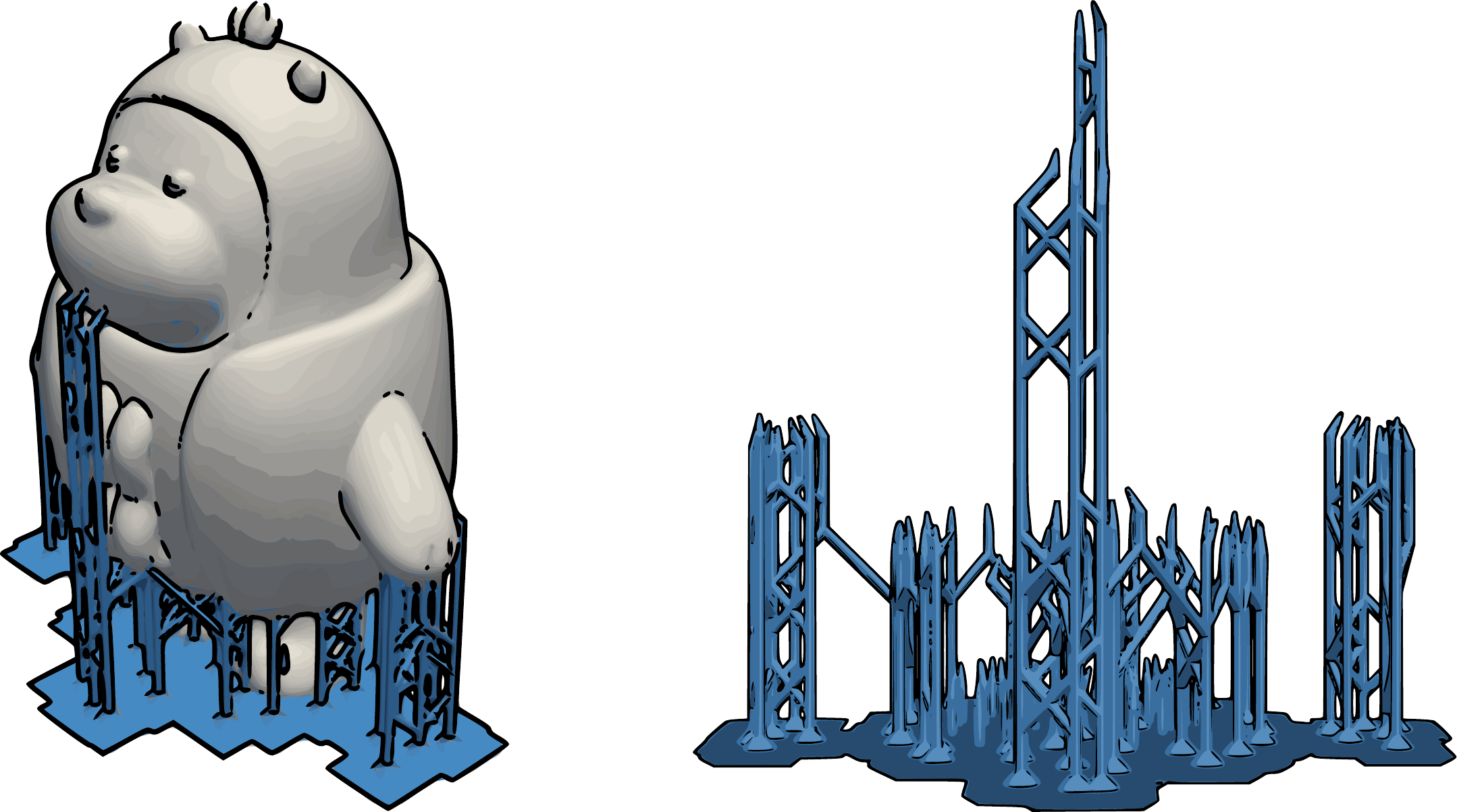}
    \caption{A model with computer-constructed geometry for 3D printing support. Since the support geometry is composed of generic primitives, local geometric words can easily over-identify otherwise unrelated models.}
    \label{fig:support}
\end{figure}

Care must be taken to deal with generic or highly common geometry. In general the frequency of geometric words is highly non-uniform, which can result in over-matching or poor quality matches. The Thangs search engine frequently encounters either: (1) generic geometric primitives, or; (2) common computer-constructed geometry. For example, algorithmically-designed support structures for 3D printing (Figure \ref{fig:support}) are present in a significant number of sculpted models.

\begin{figure}[ht]
    \centering
    \includegraphics[width=5.6in]{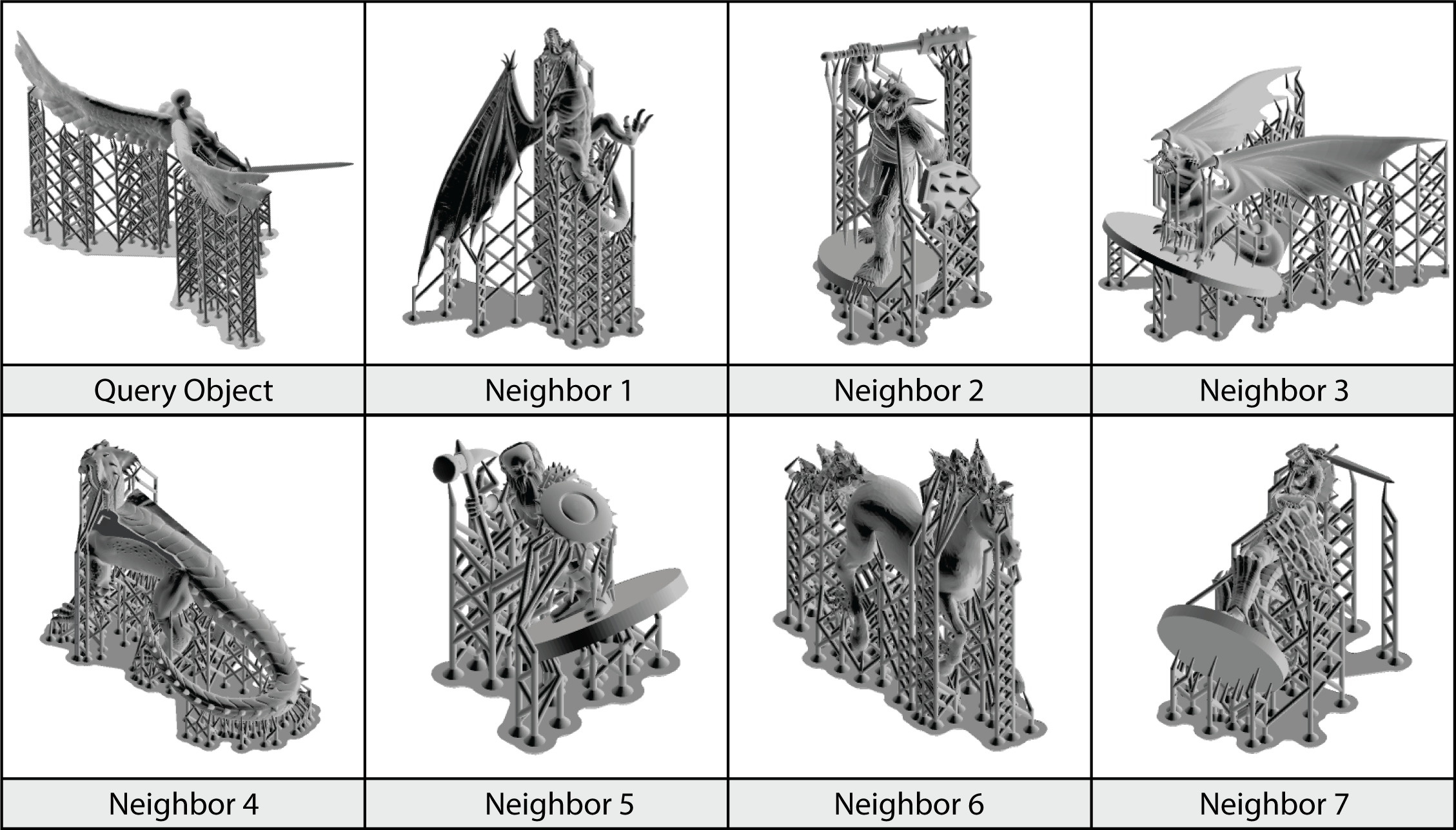}
    \caption{The neighbors of a query object using view-based 3DSS. These models would not be considered to be highly similar but for the presence of computer-constructed support geometry.}
    \label{fig:similarity_2}
\end{figure}

Computer-constructed supports are also a problem for image-based 3DSS methods. Figure \ref{fig:similarity_2} shows the nearest neighbors in a large dataset of 6 million models. The presence of support geometry leads to poor quality matches that would not occur in their absence. 
\newpage

\begin{figure}[ht]
    \centering
    \includegraphics[width=2.35in]{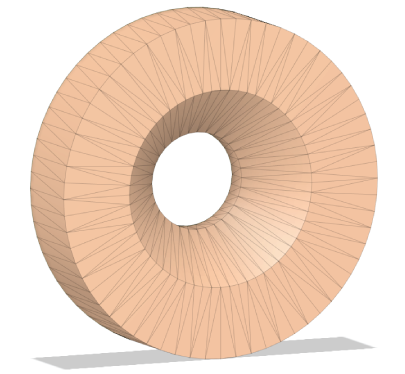}
    \caption{A CAD/CAM mesh with significant regularity. This 512-triangle model is represented by a small set of 26 local words/descriptors.}
    \label{fig:washer}
\end{figure}

CAD models are another source of common geometry. As shown in Figure \ref{fig:washer}, certain shapes can exhibit significant regularity when converted to a mesh representation. CAD models are often designed at similar scales using standardized components (e.g., M8-1.25x25mm bolts) and operations (e.g,. 45$^{\circ}$ chamfered edges). As a result, CAD models for dissimilar objects often manifest significant overlap in local geometric words; it is particularly noticeable when they have been constructed with the same CAD software and tessellated with the same library. This is a major issue in large datasets, particularly for PIP search.

\newpage
If left unchecked, the presence of highly generic geometric words in a database can result in frequent matches between seemingly unrelated models. There are several strategies that can be used to ameliorate the situation. For instance, overly generic geometric words can either be excluded from use in matching, or split into several geometric synonyms to further divide the set of shapes containing the word to more manageable bin populations.

\subsubsection{Other Sources of Collision}
\label{sec:collisions}

\begin{figure}[ht]
    \centering
    \includegraphics[width=5in]{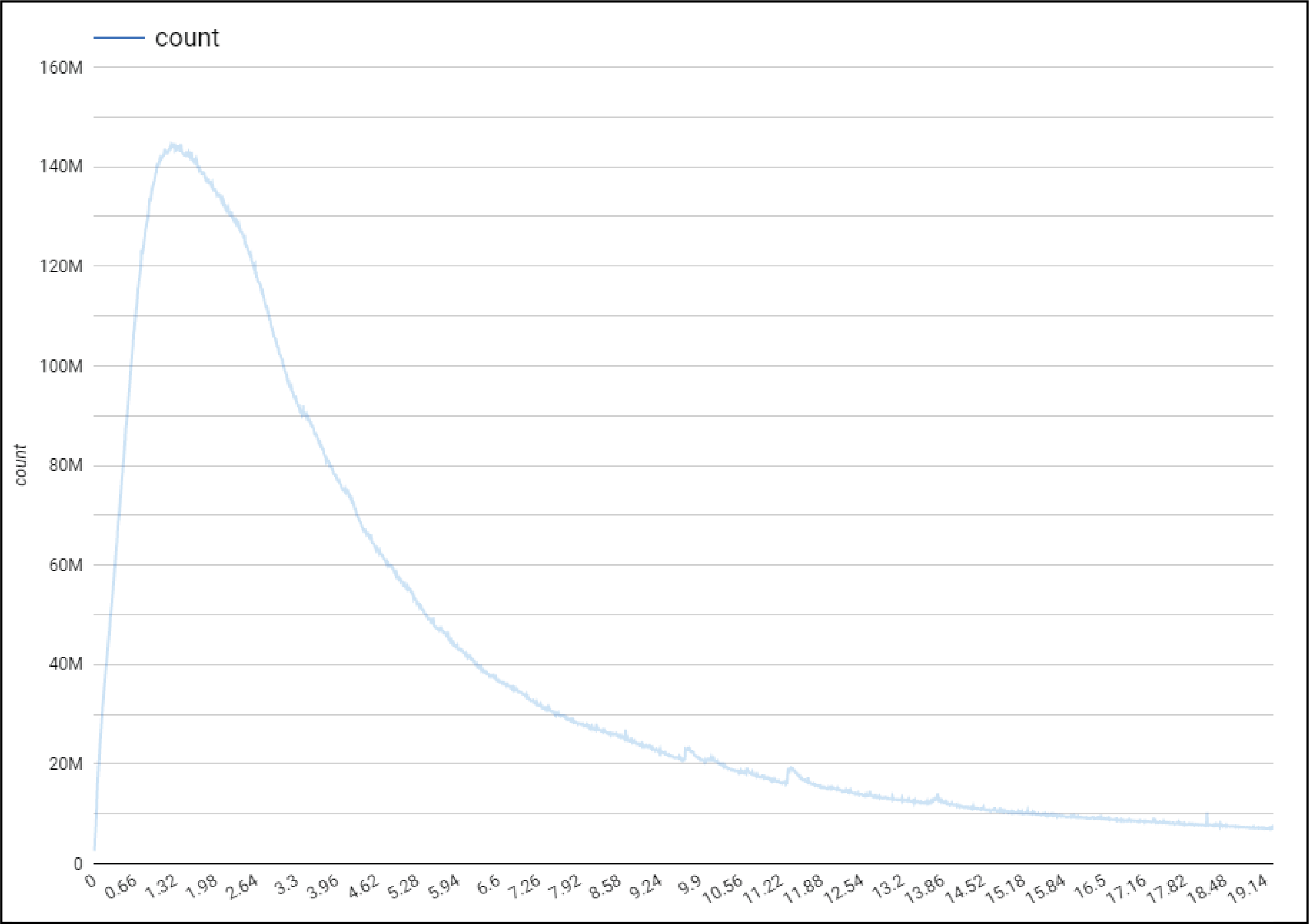}
    \caption{Triangle perimeter size across a large database. Perimeters follow a gamma distribution that clumps most of the data into a small region of space.}
    \label{fig:tridist}
\end{figure}

Even in the absence of common geometry and numerical artifacts, large databases of 3D models will invariably exhibit significant overlap in local and global geometric words. Figure \ref{fig:tridist} shows a histogram of triangle perimeter sizes for a database of 1 million 3D models. (The models are mostly sculpted, with a smaller number of CAD files). The overall distribution is gamma, showing significant degrees of overlap in the $0.5$ to $1.0$ range. Two randomly chosen bags of words are likely to have elements in common. 

This problem is not as pressing for the small (e.g., thousands of models) datasets used in research, but it becomes a major issue for the large-scale datasets that are maintained by internet-scale 3D search engines. This is particularly true for PIP search when the bag for the query object (e.g., the washer in Figure \ref{fig:washer}) is composed of only a few unique local measurements (words). In such a case, there may be numerous false positive matches if the words also happen to be prone to collision. In the Thangs dataset, it is not uncommon for certain (unfortunate) words to be found in hundreds of thousands of unrelated models.

\newpage
The issue becomes even more pronounced in the context of BOF approaches \citep{Csurka04visualcategorization,Liu06shapetopics} where vector quantization is used to map a BOW into a fixed-length feature vector (i.e., histogram). Many existing efforts in the research literature use this approach (e.g., \citep{Ohbuchi2008,Furuya2015}) with adequate performance on small datasets consisting of tens of thousands of models. On larger databases, however, the problems raised by local measurement collision are amplified by quantization.


\subsubsection{Precision Issues in Persistent Storage}

\begin{figure}[ht]
    \centering
    \includegraphics[width=6in]{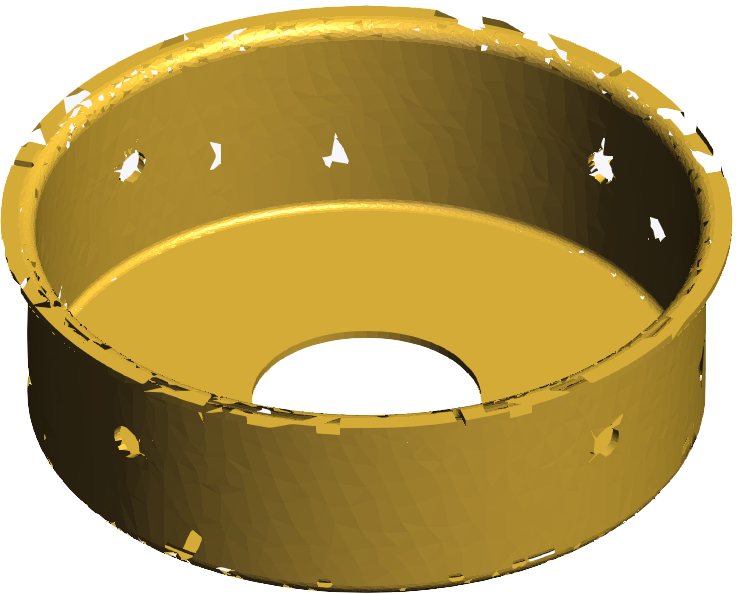}
    \caption{Artifacts from binary versus ASCII STL files. Holes in the mesh represent triangles whose local measurements fail to match between the binary and ASCII files.}
    \label{fig:file_format_1}
\end{figure}

Various precision issues arise from the manner in which 3D models are stored. For instance, file formats like STL use different numerical precision in their binary and ASCII variants. Figure \ref{fig:file_format_1} shows the intersection between binary and ASCII STL files corresponding to the same model. The gaps correspond to triangles whose local measurements (i.e., words) fail to match. In experiments, Physna has found that $1-15\%$ of triangles in an STL may be subject to this kind of numerical precision issue.
\vfill

\section{Test Datasets}

As with any IR or \textit{machine learning} (``ML'') system, robust \textit{training/test datasets} (``TTDs'') are essential for both initial development and for tracking changes (e.g., model drift) over time. Numerous open source 3D datasets exist, including: (1) Shapenet Core \citep{DBLP:journals/corr/ChangFGHHLSSSSX15}, which has  55 common object categories and  51,300 unique 3D models; (2) PartNet \citep{Mo2019PartNetAL}, which has 24 object categories and around 27,000 models; (3) the Amazon/Berkeley Object Dataset \citep{Collins_2022_CVPR}, which provides 8,0000 models, and; (4) Thingi10k, consisting of 10,000 models \citep{DBLP:journals/corr/ZhouJ16}. However, these open source datasets pale in comparison to the proprietary datasets maintained by large companies.

A good introduction to the process of generating labeled test sets is presented in \citet{Kim2020}. Among other topics, the authors discuss the steps involved in constructing a labeled dataset of roughly 60,000 CAD models. Thangs is currently supported by datasets of two types: (1) \textit{wild-type TTDs}, which contain (labeled/categorized) real-world models taken from various sources, and; (2) \textit{generated TTDs}, in which the models have been carefully selected (e.g., by applying known procedural manipulations to a set of seed models) to test properties of search algorithms. The size of Physna's TTDs varies, but on average: (1) wild-type TTDs contain hundreds of thousands of models, and; (2) generated TTDs contain tens of thousands of models. Physna also performs end-to-end regression testing on databases with model counts in the millions. 

Not only are TTDs essential for system development, but they are valuable intellectual property in their own right---in fact, they may be more valuable than patents.  Information on  training and testing methods should be considered a trade secret that must be protected by adequate safeguards, including data confidentiality policies and \textit{non-disclosure agreements} (``NDAs'').


\section{The Data Life-cycle}

Data lifecycles are a source of mundane but vexing issues for the developers of the Thangs search engine. As with many digital assets, 3D models are often in a state of evolution. For instance, authors often remove models from the internet---either retiring them or replacing them with new versions. Various practical issues will also require either the reconversion of model data or regeneration of derived data (e.g., feature vectors).

%
\subsection{Maintaining Accurate Indices}

Since 3D content is laborious to acquire, maintaining a fresh search index is a significant challenge. Web crawlers must be careful not to impose significant costs on web servers (see Section \ref{sec:crawling_regulation}). Unfortunately, 3D assets can be quite large; even worse, some 3D websites handle download requests by generating (rather than serving) files---that is, they extract a set of parameters from an HTTP request and invoke software components that generate a model for download. This type of website is not designed for high throughput and can be heavily impacted by even moderate \textit{queries per second} (``QPS'').

As a result of these and other issues, a 3D search engine faces significant challenges in maintaining an accurate inventory of models available on the web. In contrast to text-based search engines, it is not possible to guarantee freshness of 3D assets by recrawling websites on a frequent (e.g,. daily) basis. The Thangs search engine deals with this issue by maintaining freshness estimates in the database, while its crawlers are bifurcated into model crawlers and metadata crawlers. The latter are tasked with daily text mining of site metadata; if they find that a webpage with 3D content has been updated, it is added to a recrawl queue. Recrawls are attempted on a per-site schedule that takes into account the target site's \textit{robots.txt} file, server capacity, and other factors. Models that are no longer available to the public are taken offline in an automated batch job.

%
\subsection{Preview Images}

3D search engine operators should be aware that preview images are of significant importance to content creators. The initial preview images shown in the Thangs search engine were rendered directly from the canonical database of STL files. Physna received numerous complaints from both Thangs content creators and OEM website operators that the preview images were not a flattering representation of the web pages that they pointed to. The Thangs team had to build custom workflows that allow Thangs content creators to choose their own preview images. RESTful \textit{application programming interfaces} (``APIs'') were introduced so that OEM website operators could also submit recommended preview images for a given model. 

This is, of course, a significant deviation from the types of dataflows found within a typical search engine (e.g., Google, Bing, Yandex). As detailed in Section \ref{sec:data_sources}, a large number of internet-accessible 3D models are found on proprietary websites operated by OEMs who typically restrict access to content through registration mechanisms and TOU documents. These website operators are in a position to make feature requests in exchange for granting access to their datasets.

\subsection{Re-generation of Data}

There is a significant amount of rework involved in maintaining an  online 3D search engine. First, as discussed in Section \ref{sec:file_formats}, the use of a single canonical file format means that the entire database is dependent on a particular version of a conversion library; if the library version changes, the various model files may need to be reconverted/regenerated to ensure consistency. Second, information retrieval schemes invariably use secondary representations, such as: (1) multisets of local measurements, for BOW approaches, or; (2) rendered images, for view-based approaches. These secondary representations must be regenerated if there is a change in the basic algorithm or a defect that needs to be resolved. The cost involved in regenerating information for tens of millions of 3D models is not trivial.

\section{Conclusion}

This paper has discussed the major data management challenges facing internet-scale 3D search engines, including model acquisition, diversity of file formats, shape retrieval, intellectual property, the legality of web crawling, and trustworthy computing. While some of these issues, such as privacy and security, are shared with other information systems, others are unique to this domain. The discussion contains some guidance on how Physna has approached some of these challenges, but a thorough treatment is well beyond  a single paper.

Apart from ongoing work on improving 3D search methods there are numerous areas of research that directly impact the challenges mentioned above. First, researchers should continue to investigate the use of 3D model files as attack vectors on 3D search engines. While this is largely a theoretical concern at present, if 3D search engines become highly visible they will inevitably attract attention from malicious actors. In particular, this type of attack may be appealing to competitors and other actors skilled at corporate espionage.

Second, further research on the detection of objectionable (e.g., obscene) shapes is sorely needed. There is some existing work that converts 3D objects to 2D images and then uses 2D imaging techniques to classify the results. However, from an organizational standpoint it would be useful to find techniques that operate within the 3D domain directly. In particular it is difficult for startups to develop and maintain multiple systems operating in parallel.

Third, relatively little work in the academic literature has addressed information retrieval for internet-scale 3D model collections (including means for indexing large numbers of BOW representations).  In contrast to the relatively small datasets found in the research literature, datasets for an internet-scale 3D search engine will contain tens of millions of models at minimum. Certain issues become more salient when the size of the dataset is extremely large. For example, local neighborhood estimates start to overlap in  datasets of this size, causing serious problems for PIP search.

Fourth, there is a clear need for further research on remixing of 3D models. In particular, the field lacks detailed ethnographic research in which user attitudes towards remixing are elicited and analyzed. At least one previous work \citep{Flath2017} has also identified this gap. 3D content creation platforms like Thingiverse are hotspots of creativity, and remixing seems to be a significant design strategy. However, balancing the interests of commercial and open source users is vital for search engines that need to gain access to the vast store of proprietary models on the internet.

\section*{Acknowledgments}

The authors would like to thank Physna's management for granting permission to draw upon operational data. This work was written by authors Williams and Scott, with Wedig, Hindanov, and Roedig contributing various insights, data and diagrams.

\newpage
\bibliography{references}
\end{document}